\documentstyle[epsf,graphicx,amssymb,amsfonts,amsmath,mathptm]{mn}

\newif\ifAMStwofonts \AMStwofontstrue 
\DeclareMathVersion{bold} 
\DeclareMathAlphabet{\mathbfit}{OT1}{cmr}{bx}{it}
\SetMathAlphabet\mathbfit{bold}{OT1}{cmr}{bx}{it}
\DeclareMathAlphabet{\mathbfss}{OT1}{cmss}{bx}{n}
\SetMathAlphabet\mathbfss{bold}{OT1}{cmss}{bx}{n}
\ifAMStwofonts 
  \DeclareSymbolFont{UPM}{U}{eur}{m}{n}
  \SetSymbolFont{UPM}{bold}{U}{eur}{b}{n}
  \DeclareSymbolFont{AMSa}{U}{msa}{m}{n}
  \DeclareMathSymbol{\upi}{0}{UPM}{"19}
  \DeclareMathSymbol{\umu}{0}{UPM}{"16}
  \DeclareMathSymbol{\upartial}{0}{UPM}{"40}
  \DeclareMathSymbol{\leqslant}{3}{AMSa}{"36}
  \DeclareMathSymbol{\geqslant}{3}{AMSa}{"3E} 
     \let\leq=\leqslant
   \let\geq=\geqslant  

\title [CMB for a Nearly Flat Compact Hyperbolic Universe] {The Cosmic
Microwave Background for a Nearly Flat Compact Hyperbolic Universe}
\author[R.~Aurich and F.~Steiner] {R.~Aurich and F.~Steiner\\
Abteilung Theoretische Physik, Universit\"at Ulm\\
Albert-Einstein-Allee 11, D-89069 Ulm \\ Federal Republic of Germany }

\date{Accepted ???. Received ???; in original form \today}
\pagerange{\pageref{firstpage}--\pageref{lastpage}} \pubyear{2000}
\begin{document}
\maketitle
\label{firstpage}

\begin{abstract}
The fluctuations of the cosmic microwave background (CMB) are
investigated for a hyperbolic universe with finite volume.
Four-component models with radiation, matter, vacuum energy, and an
extra spatially constant dark energy $X$-component are considered.
The general solution of the Friedmann equation for the cosmic scale factor
$a(\eta)$ is given for the four-component models in terms of the
Weierstrass ${\cal P}$-function.
The lower part of the angular power spectra $C_l$ of the CMB anisotropy
is computed for nearly flat models with
$\Omega_{\hbox{\scriptsize{tot}}}\leq 0.95$.
It is shown that the particular compact fundamental cell, which is
considered in this paper, leads to a suppression in $C_l$ for
$l\lesssim 10$ and $\Omega_{\hbox{\scriptsize{tot}}} \lesssim 0.9$.
\end{abstract}


\section{Introduction}

The two crucial properties of the universe at large scales are its
curvature and its topology.  Both properties are encoded in the cosmic
microwave background (CMB), see e.\,g.\
\cite{Hu_White_1996,Cornish_Spergel_Starkman_1998,Kamionkowski_Buchalter_2000},
which is measured with ever increasing resolution.  The detection of
the first acoustic peak in the CMB angular power spectrum $C_l$
\cite{Knox_Page_2000} provides evidence for a flat, or nearly flat,
cold dark matter universe with a non-vanishing cosmological constant
and/or an extra exotic energy component.  The recent Boomerang
\cite{deBernardis_et_al_2000} and MAXIMA-1 \cite{Hanany_et_al_2000}
measurements yield evidence even for the second acoustic peak which
is, however, less pronounced than expected by standard CMB models.
These CMB scenarios are based on isentropic initial perturbations in a
universe composed of radiation, baryonic matter according to the Big
Bang nucleosynthesis, cold dark matter, and a non-vanishing
cosmological constant.  Possible explanations for the low second peak
and the surprisingly large scale of the first peak are discussed in
\cite{White_Scott_Pierpaoli_2000,%
Tegmark_Zaldarriaga_2000,Cornish_2000,Weinberg_2000b}.  The
constraints obtained from the MAXIMA-1 experiment are
$\Omega_{\text{tot}}=0.90\pm 0.15$, $\Omega_{\text{bar}}h_0^2=0.025\pm
0.010$, $\Omega_{\text{cdm}}h_0^2=0.13\pm 0.10$, and a spectral index
$n=0.99\pm 0.09$ at 95\% confidence level \cite{Balbi_et_al_2000}.
Here $\Omega_{\text{cdm}}$ and $\Omega_{\text{bar}}$ denote the ratio
of the cold dark matter (cdm) and baryonic (bar) energy densities,
respectively, to the critical energy density.

The standard models describe the structure of the acoustic peaks, but
fail to match with the low quadrupole moment $C_2$ of the COBE
experiment \cite{Tegmark_Hamilton_1997}.  This can be interpreted as a
hint for a non-trivial topology of our universe.  The standard models
suppose a trivial topology implying a universe with infinite volume
for negative and zero curvature.  Models with a non-trivial topology
lead to a finite volume and to a suppression in the angular power
spectrum $C_l$ for low multipoles.  This motivates the study of
non-trivial topologies \cite{Lachieze-Rey_Luminet_1995} which leads to
multiple images of a single source called topological lensing
\cite{Uzan_Lehoucq_Luminet_2000}, and the just mentioned suppression
in the large scale CMB anisotropy on which we concentrate in this
paper.  For flat models the topological length scale is constrained to
be significantly larger than half the diameter of the observable
universe \cite{Levin_Scannapieco_Silk_1998} which renders these models
unattractive.  (See, however, \cite{Roukema_2000}.)  Thus in the
following we discuss models with negative curvature.

The computation of the CMB anisotropy for compact hyperbolic universes
can be carried out in two different ways.  On the one hand one can
compute the fluctuations by using the so--called method of images
which requires the group elements which define the fundamental cell of
the considered non-trivial topology
\cite{Bond_Pogosyan_Souradeep_1998,Bond_Pogosyan_Souradeep_1999a,%
Bond_Pogosyan_Souradeep_1999b}.  On the other hand one can use a
method which requires the eigenmodes of the fundamental cell with
respect to the Laplace-Beltrami operator of the considered space.
With the latter method the CMB anisotropy is computed for hyperbolic
universes with a vanishing cosmological constant $\Lambda$ for a
compact orbifold \cite{Aurich_1999} and several compact manifolds
\cite{Inoue_Tomita_Sugiyama_1999,Cornish_Spergel_1999}.

Our main aim in this paper is to incorporate a non-vanishing
cosmological constant $\Lambda$ and, in addition, an extra smooth
dark energy component $\varepsilon_{\text{x}}$ \cite{Turner_White_1997}
in the anisotropy calculations as suggested by current observations,
where we concentrate on the lower part of the angular power spectrum $C_l$,
which is affected by the non-trivial topology.
Recent investigations, in particular of the luminosity-redshift surveys
of Type Ia supernovae
\cite{Riess_et_al_1998,Perlmutter_et_al_1999}, strongly indicate that
current observations require apart from a matter density component
$\Omega_{\text{mat}}=\Omega_{\text{cdm}}+\Omega_{\text{bar}}\simeq0.3\pm
0.1$ an additional unclustered, dark energy component of 60\% of the
total energy density of the universe with negative pressure
\cite{Wang_Cladwell_Ostriker_Steinhardt_2000,Hu_Fukugita_Zaldarriaga_Tegmark_2000}
corresponding to an accelerated expansion.  In the following we will
assume that the new component is a mixture of vacuum energy,
$\Omega_{\text{vac}}$, or cosmological constant $\Lambda$, and a smooth
dark energy component $\Omega_{\text{x}}$.
Whereas a cosmological constant $\Lambda>0$
corresponds to a constant homogeneous energy component
$\varepsilon_{\text{vac}}>0$ with negative pressure and equation of
state $w_{\text{vac}}=-1$, the extra dark energy component $X$
considered by us consists of a dynamical, time-dependent energy
density $\varepsilon_{\text{x}}>0$ with negative pressure and
equation of state $w_{\text{x}}=-\frac 23$.  (Here $w$ denotes the
ratio of pressure to energy density.)
The $X$-component is similar to a particular version of quintessence
which is generated by a slowly evolving scalar field
with an exponential or inverse power law potential and $-1 <
w_{\text{quint}} \leq 0$ \cite{Ratra_Peebles_1988,Peebles_Ratra_1988,%
Wetterich_1988_a,Wetterich_1988_b,Caldwell_Dave_Steinhardt_1998}.  The
quintessence models and the more recent ``tracker field'' models
\cite{Zlatev_Wang_Steinhardt_1999} have been introduced to solve the
two cosmological constant problems \cite{Weinberg_2000a}: i) why is
$\varepsilon_{\text{vac}}$ so small, ii) why is
$\varepsilon_{\text{vac}}$ not only small, but also of the same
magnitude as the present mass of the universe?  (See also
\cite{Armendariz-Picon_Mukhanov_Steinhardt_2000a,%
Armendariz-Picon_Mukhanov_Steinhardt_2000b} for the concept of
``k-essence'' and \cite{Weinberg_2000a} for an anthropic explanation.)
While at present there is no direct evidence for a scalar field, let
alone for a particular form of the potential, observations are
consistent with an $X$-component if $w_{\text{x}}=-0.65\pm0.07$
(assuming flat models) \cite{Wang_Cladwell_Ostriker_Steinhardt_2000}.
However, since the concept of a scalar quintessence field is a purely
classical phenomenological one, we expect the quantum mechanics of
such a theory to be plagued with the usual problem of
nonrenormalizability \cite{Ratra_Peebles_1988,Peebles_Ratra_1988}.  We
therefore adopt in this paper the point of view of an effective
quintessence model, where we do not start from a given potential for
the scalar field, but rather give the redshift behavior
$\varepsilon_{\text{x}} \sim a^{-1}$ and thus the equation of
state $w_{\text{x}}=-\frac 23$, which is consistent with the above
cited experimental bounds ($a$ is the cosmic scale factor, see the
next section).  Furthermore, we will assume that the $X$-component
is spatially constant which can be understood as follows
\cite{Ratra_Peebles_1988,Peebles_Ratra_1988}.  In linear perturbation
theory spatial gradients in the scalar field act like particles with
very low mass that cannot bind to a nonrelativistic gravitational
potential well that is much smaller than the Hubble length $H^{-1}$.
Thus dynamical studies of groups and clusters of galaxies with size
$\ll H^{-1}$ cannot detect concentrations in
$\varepsilon_{\text{x}}$, and thus $\varepsilon_{\text{x}}$
can be assumed to be spatially constant.
This is confirmed by recent determinations of $\Omega_{\text{mat}}$
obtained from relative velocities of galaxies yielding results
in the range $0.2 \lesssim \Omega_{\text{mat}} \lesssim 0.4$
\cite{Juszkiewicz_Ferreira_Feldman_Jaffe_Davis_2000,Willick_2000}.
Since these measurements are sensitive to a spatially inhomogeneous dark
energy component \cite{Lahav_Lilje_Primack_Rees_1991},
we have to assume the dark energy component,
required by CMB measurements, to be smooth.
Therefore, we assume below an $X$-component being spatially
homogeneous like the vacuum energy and $\Omega_{\text{mat}}=0.3$.
(For further arguments concerning the smoothness assumption,
see \cite{Turner_White_1997,White_1998}.)

The fundamental cell which we consider in this paper is the same
pentahedron as considered in \cite{Aurich_1999} with the same
Dirichlet eigenmode spectrum.  (For more details, see also
\cite{Aurich_Marklof_1996}.)  One is interested in fundamental cells
with volumes as small as possible for several reasons, e.\,g.\ the
creation probability of the universe increases dramatically with
decreasing volume \cite{Atkatz_Pagels_1982} and, furthermore, to
obtain an appreciable effect in the CMB anisotropy.  Unfortunately,
the smallest hyperbolic manifold ${\cal M}$ is unknown.  The smallest
one known has a volume $\hbox{vol}({\cal M}) = 0.94272\dots R^3$,
which is far above the lower bound $\hbox{vol}({\cal M}) >
0.16668\dots R^3$ \cite{Gabai_Meyerhoff_Thurston_1996}.  Here $R$
denotes the curvature radius of the universal covering space.

Concerning the impact on the CMB anisotropy the main difference with
respect to infinite-volume models arises from the discrete eigenvalue
spectrum $\{E_n\}$ which compact manifolds possess in contrast to the
continuous spectrum of the infinite volume models.  The volume of the
manifold has an important influence on the CMB anisotropy because it
determines the magnitude of the lowest eigenmode and the number of
eigenmodes below a given value $E$.  For a manifold ${\cal M}$, the
number ${\cal N}(E)$ of eigenvalues below $E$ is asymptotically given
by Weyl's law $(R=1)$
$$
{\cal N}(E) \; \sim \; \frac{\hbox{vol}({\cal M})}{6\pi^2} \, k^3
\hspace{10pt} \hbox{with} \hspace{10pt} k := \sqrt{E-1}
\hspace{10pt} .
$$
The first lowest eigenmodes determine the largest scales of the CMB
anisotropies.  The smaller the volume the stronger is the suppression
of the first multipoles in the angular power spectrum $C_l$.

\begin{figure}
\vspace*{-15pt}
\begin{center}
\hspace*{-25pt}\includegraphics[width=9cm]{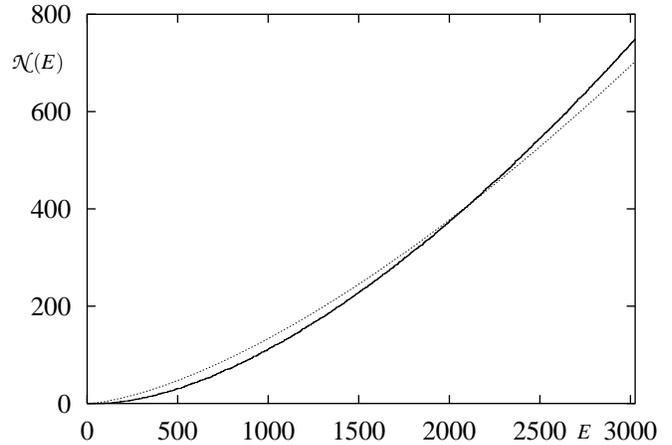}
\put(-28,12){$E$} \put(-242,152){${\cal N}(E)$}
\end{center}
\vspace*{-15pt}
\caption{\label{Fig:N_von_E} The number ${\cal N}(E)$ of eigenvalues
below $E$ is shown together with Weyl's law for the tetrahedral
fundamental cell with Dirichlet boundary conditions (dashed curve, not
visible) and Weyl's law for a manifold with volume 0.25 (dotted curve),
$R=1$.  }
\end{figure}

The considered pentahedron has a volume $\hbox{vol}({\cal M}) \simeq
0.7173068 R^3$.  Since only one of two symmetry classes is taken into
account, the following computations correspond to a hyperbolic
manifold with $\hbox{vol}({\cal M}) \simeq 0.3586534 R^3$.  However, a
volume comparison is complicated by the fact that Weyl's law has
additional terms for orbifolds in comparison to manifolds, in
particular the surface term \cite{Aurich_Marklof_1996} being absent in
the case of manifolds.  The additional surface term leads to a
suppression of ${\cal N}(E)$ in comparison with manifolds as shown in
figure \ref{Fig:N_von_E} for $E<3026$.  The figure demonstrates that
the considered orbifold mimics a manifold with effective volume
$\simeq 0.25 R^3$.  The statistical properties of the eigenmodes are
expected to be of the same random nature as observed in quantum chaos
\cite{Aurich_Steiner_1993,Inoue_1999}.

\section{The background model}

The standard cosmological model based on the
Fried\-mann-Le\-ma\^{\i}tre-Robertson-Walker metric $(c=1)$
$$
ds^2 \; = \; a^2(\eta) \left\{ d\eta^2 - \gamma_{ij} dx^i dx^j
\right\}
$$
is governed for negative curvature ($K=-1$) by the Friedmann equation
$$
{a'}^2 - a^2 \; = \; \frac{8\pi G}3 T_0^0 a^4
\hspace{10pt} ,
$$
where $a(\eta)$ is the cosmic scale factor and $\eta$ the conformal time.
The prime denotes differentiation with respect to $\eta$.  The
energy-momentum tensor for an ideal fluid is given by
$$
T^\mu_\nu \; = \; (\varepsilon + p) u^\mu u_\nu \, - \, p \,
\delta^\mu_\nu
\hspace{10pt} ,
$$
where $u^\mu$ is the four-velocity of the fluid, and
$\varepsilon=\varepsilon(\eta)$ denotes the energy density and
$p=p(\eta)$ the pressure.  In the following we consider
multi-component models containing a matter-energy density
$\varepsilon_{\text{mat}}$, a radiation density
$\varepsilon_{\text{rad}}$ as well as a non-vanishing cosmological
constant $\Lambda=8\pi G\varepsilon_{\text{vac}}$ and a spatially
constant $X$-component $\varepsilon_{\text{x}}$ with an
equation of state $p_{\text{x}}=-\frac 23
\varepsilon_{\text{x}}$.  Then the $00-$component of the
energy-momentum tensor is given in comoving coordinates by
$$
T_0^0 \; = \; \sum_{\substack{k=0 \\ k \neq 2}}^4 \varepsilon_{k,0}
\left( \frac{a_0}a \right)^k
\hspace{10pt} ,
$$
expressed in terms of the current radiation density $\varepsilon_{4,0}
= \varepsilon_{\text{rad},0}$, the current matter density
$\varepsilon_{3,0} = \varepsilon_{\text{mat},0}$, the current
$X$-component energy density $\varepsilon_{1,0} =
\varepsilon_{\text{x},0}$ and a vacuum energy density
$\varepsilon_{0,0} = \varepsilon_{\text{vac}}$.  Here $a_0 :=
a(\eta_0)$ is the scale factor of the present epoch.  The present
conformal time $\eta_0$ is implicitly given by
$$
a(\eta_0) \, = \, \frac 1{H_0 \sqrt{1-\Omega_{\text{tot}}}}
\hspace{5pt} , \hspace{5pt} \Omega_{\text{tot}} = \Omega_{\text{rad}}
+ \Omega_{\text{mat}} + \Omega_{\text{x}} + \Omega_{\text{vac}}
\hspace{5pt} ,
$$
where $H_0 = h_0 100 \hbox{ km s}^{-1} \hbox{Mpc}^{-1}$ denotes
Hubble's constant and $\Omega_k :=
\varepsilon_{k,0}/\varepsilon_{\text{crit}}$ with
$\varepsilon_{\text{crit}} = 3 H_0^2 /(8\pi G)$.  With
$$
\Omega_2 := \Omega_{\text{curv}} := -\frac K{(a_0 H_0)^2} = 
\frac 1{(a_0 H_0)^2} = 1 - \Omega_{\text{tot}} > 0 
$$
the Friedmann equation reads
\begin{equation}
\label{Eq:Friedmann_three_component}
a'(\eta) \; = \; H_0 \, \sqrt{ \sum_{k=0}^4 \Omega_k a_0^k a^{4-k} }
\hspace{5pt} .
\end{equation}
This gives the infinitely far future $\eta_\infty$ as
\begin{equation}
\label{Eq:eta_infty}
\eta_\infty \; = \; \sqrt{1 - \Omega_{\text{tot}}} \,
\int_0^\infty \frac{dx}{\sqrt{\sum_{k=0}^4 \Omega_k x^k} }
\hspace{10pt} ,
\end{equation}
which yields $\eta_\infty < \infty$ for a large class of models, see
below.  Notice that the various components redshift like $a^{-k}$ with
an associated equation of state $w_k := p_k/\varepsilon_k = (k-3)/3$.
The current value of the deceleration parameter is given by
$q_0 = \Omega_{\text{rad}} + \frac 12 \Omega_{\text{mat}}
- \frac 12 \Omega_{\text{x}} - \Omega_{\text{vac}}$.

Let us define the following quantities
$$
A \; := \; \frac 12 \Omega_{\text{mat}} H_0^2 a_0^3 \; = \; \frac{2
a_{\text{eq}}}{\hat{\eta}^2}
\hspace{10pt} ,
$$
$$
B \; := \; \frac 14 \Omega_{\text{x}} H_0^2 a_0
\hspace{10pt} ,
$$
$$
C \; := \; \frac 1{12} A^2 \hat\eta^2 \Lambda
$$
with
$$
\hat{\eta} \; := \;
\frac{2\sqrt{\Omega_{\text{rad}}}}{H_0 a_0 \Omega_{\text{mat}}} \;
\simeq \; \left( 1 + \sqrt 2\, \right) \, \eta_{\text{eq}}
\hspace{10pt} ,
$$
where the subscript ``eq'' marks the epoch of matter-radiation
equality, and $a_{\text{eq}} := a(\eta_{\text{eq}}) = a_0
(\Omega_{\text{rad}}/\Omega_{\text{mat}})$.  With the initial
conditions $a(0)=0$ and $a'(0)>0$, equation
(\ref{Eq:Friedmann_three_component}) has the unique solution
\begin{equation}
\label{Eq:Scale_Factor}
a(\eta) \; = \; \frac{A}{2} \, \frac{ {\cal P}(\eta) - \frac 1{12} -
\hat{\eta}\, {\cal P}'(\eta) + AB\hat\eta^2} {({\cal P}(\eta) - \frac
1{12})^2 \, - \, C}
\hspace{10pt} .
\end{equation}
Here ${\cal P}(\eta)$ denotes the Weierstrass ${\cal P}$-function
which can numerically be evaluated very efficiently by
\begin{equation}
\label{Eq:Weierstrass}
{\cal P}(\eta) \; = \; {\cal P}(\eta;g_2,g_3) \; = \; \frac 1{\eta^2}
\, + \, \sum_{k=2}^\infty c_k \eta^{2k-2}
\end{equation}
with
$$
c_2 := \frac{g_2}{20}
\hspace{10pt} , \hspace{10pt} c_3 := \frac{g_3}{28}
$$
and \cite{Abramowitz_Stegun_1965}
$$
c_k = \frac{3}{(2k+1)(k-3)} \sum_{m=2}^{k-2} c_m c_{k-m} \; \hbox{ for
} \; k\geq 4
\hspace{10pt} .
$$
The so-called invariants $g_2$ and $g_3$ are determined by the
cosmological parameters
\begin{eqnarray} \nonumber
\hspace{15pt} g_2 & = & \frac 1{12} + 4 C - 2 A B \\ \nonumber
\hspace{15pt} g_3 & = & -\,\frac 1{216} + \frac{8C-A^2\Lambda}{12} +
\frac{AB}6 - A^2 B^2 \hat\eta^2
\hspace{10pt} .
\end{eqnarray}
For cosmologically plausible parameter choices the series
(\ref{Eq:Weierstrass}) needs only to be evaluated by taking into
account the first twenty terms and thus the explicit solution
(\ref{Eq:Scale_Factor}) is much more efficient than the usual
integration of the Friedmann equation.

Expanding (\ref{Eq:Scale_Factor}) in a series at $\eta=0$ gives the
scale factor at early times
\begin{eqnarray}\nonumber
\hspace{10pt} a(\eta) & = & A \, \left\{ \hat\eta\,\eta +
\frac{\eta^{2}}2 + \hat\eta\,\frac{\eta^3}{3!} + \left( 1 + 12\,A
B{\hat\eta}^{2} \right )\frac{\eta^4}{4!} \right. \\ & & \nonumber
\left. \hspace{30pt} + \hat\eta\,\left( 1 + 36 A B + 48 C \right
)\frac{\eta^5}{5!}  +O\left ({\eta}^{6}\right) \right\}
\hspace{10pt} .
\end{eqnarray}
This expansion shows that the $X$-component term $B$ influences the
scale factor one power in $\eta$ lower than the cosmological constant
term $C$.

\begin{figure}
\vspace*{-15pt}
\begin{center}
\hspace*{-15pt}\includegraphics[width=9cm]{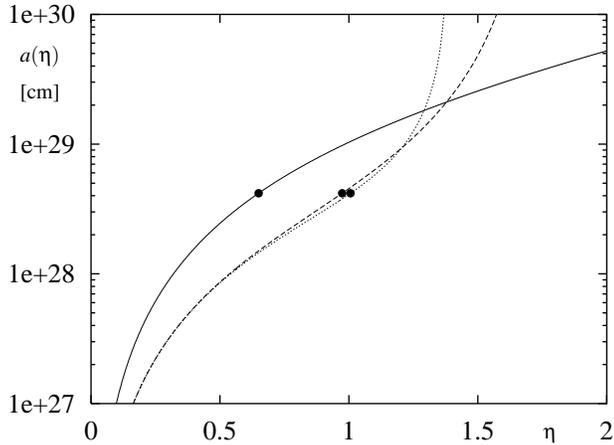}
\put(-30,12){$\eta$} \put(-228,155){$a(\eta)$} \put(-228,141){[cm]}
\end{center}
\vspace*{-15pt}
\caption{\label{Fig:scale_factor} The scale factor is shown for nearly
flat models with $\Omega_{\text{tot}} = 0.9$ and $h_0=0.7$.  The full
curve corresponds to $\Omega_{\text{mat}} = 0.9$,
$\Omega_{\text{x}} = \Omega_{\text{vac}} = 0.0$, the dashed curve
to $\Omega_{\text{mat}} = 0.3$, $\Omega_{\text{x}} = 0.6$ and
$\Omega_{\text{vac}} = 0.0$, and the dotted curve to
$\Omega_{\text{mat}} = 0.3$, $\Omega_{\text{x}} = 0.0$ and
$\Omega_{\text{vac}} = 0.6$.  The last model has the smallest
$\eta_\infty$.  The present scale factor $a_0$ is indicated by a dot.
}
\end{figure}

\begin{table*}
\centerline{
\begin{tabular}{|ccccccccc|}
\hline $\Omega_{\text{x}}$ & $\Omega_{\text{vac}}$ & $a_0$ [cm] &
$t_0$ [Gyr] & $\eta_{\text{eq}}$ & $\eta_{\hbox{\scriptsize{SLS}}}$ &
$\eta_0$ & $\eta_\infty$ & $N_{\text{Pentahedron}}$ \\ \hline
0.00 & 0.00 & 15.797E+27 &  11.30 &  0.016 &  0.057 &  2.381 & $\infty$ &  208 \\
0.10 & 0.00 & 17.063E+27 &  11.46 &  0.015 &  0.053 &  2.231 &  5.364 &  151 \\
0.20 & 0.00 & 18.692E+27 &  11.62 &  0.014 &  0.048 &  2.062 &  4.444 &  105 \\
0.30 & 0.00 & 20.899E+27 &  11.80 &  0.012 &  0.043 &  1.868 &  3.765 &   68 \\
0.40 & 0.00 & 24.132E+27 &  11.98 &  0.011 &  0.037 &  1.639 &  3.146 &   39 \\
0.50 & 0.00 & 29.557E+27 &  12.18 &  0.009 &  0.030 &  1.357 &  2.505 &   19 \\
0.60 & 0.00 & 41.805E+27 &  12.39 &  0.006 &  0.022 &  0.974 &  1.739 &    6 \\
\hline
0.00 & 0.10 & 17.063E+27 &  11.53 &  0.015 &  0.053 &  2.240 &  3.780 &  154 \\
0.10 & 0.10 & 18.692E+27 &  11.70 &  0.014 &  0.048 &  2.070 &  3.424 &  107 \\
0.20 & 0.10 & 20.899E+27 &  11.88 &  0.012 &  0.043 &  1.876 &  3.047 &   69 \\
0.30 & 0.10 & 24.132E+27 &  12.08 &  0.011 &  0.037 &  1.647 &  2.630 &   40 \\
0.40 & 0.10 & 29.557E+27 &  12.28 &  0.009 &  0.030 &  1.364 &  2.144 &   19 \\
0.50 & 0.10 & 41.805E+27 &  12.49 &  0.006 &  0.022 &  0.979 &  1.516 &    6 \\
\hline
0.00 & 0.20 & 18.692E+27 &  11.79 &  0.014 &  0.048 &  2.079 &  3.253 &  109 \\
0.10 & 0.20 & 20.899E+27 &  11.97 &  0.012 &  0.043 &  1.885 &  2.909 &   70 \\
0.20 & 0.20 & 24.132E+27 &  12.17 &  0.011 &  0.037 &  1.655 &  2.521 &   41 \\
0.30 & 0.20 & 29.557E+27 &  12.38 &  0.009 &  0.030 &  1.371 &  2.063 &   20 \\
0.40 & 0.20 & 41.805E+27 &  12.60 &  0.006 &  0.022 &  0.984 &  1.463 &    6 \\
\hline
0.00 & 0.30 & 20.899E+27 &  12.07 &  0.012 &  0.043 &  1.893 &  2.830 &   72 \\
0.10 & 0.30 & 24.132E+27 &  12.27 &  0.011 &  0.037 &  1.663 &  2.458 &   42 \\
0.20 & 0.30 & 29.557E+27 &  12.48 &  0.009 &  0.030 &  1.378 &  2.015 &   20 \\
0.30 & 0.30 & 41.805E+27 &  12.71 &  0.006 &  0.022 &  0.989 &  1.432 &    6 \\
\hline
0.00 & 0.40 & 24.132E+27 &  12.37 &  0.011 &  0.037 &  1.671 &  2.416 &   43 \\
0.10 & 0.40 & 29.557E+27 &  12.59 &  0.009 &  0.030 &  1.385 &  1.983 &   20 \\
0.20 & 0.40 & 41.805E+27 &  12.82 &  0.006 &  0.022 &  0.995 &  1.411 &    6 \\
\hline
0.00 & 0.50 & 29.557E+27 &  12.70 &  0.009 &  0.030 &  1.392 &  1.960 &   21 \\
0.10 & 0.50 & 41.805E+27 &  12.94 &  0.006 &  0.022 &  1.000 &  1.396 &    6 \\
\hline
0.00 & 0.60 & 41.805E+27 &  13.06 &  0.006 &  0.022 &  1.006 &  1.385 &    6 \\
\hline
\end{tabular}
}
\caption{ \label{Tabelle} The present value of the scale factor $a_0$
and the present age of the universe $t_0$ as well as several important
conformal times, i.\,e.\ $\eta_{\text{eq}}$,
the time of recombination $\eta_{\hbox{\scriptsize{SLS}}}$ at redshift $z=1200$,
$\eta_0$ and $\eta_\infty$, are
given for $\Omega_{\text{mat}}=0.3$ and $h_0=0.7$ and different
combinations of $\Omega_{\text{x}}$ and $\Omega_{\text{vac}}$.
The number $N_{\text{Pentahedron}}$ of pentahedrons within the surface
of last scattering is also presented.  }
\end{table*}

In the case $\Lambda > 0$ and/or $B>0$ the conformal time $\eta$ is
restricted to $0 \leq \eta < \eta_\infty < \infty$, where
$\eta_\infty$, defined in (\ref{Eq:eta_infty}), is obtained from the
implicit relation
$$
{\cal P}(\eta_\infty;g_2,g_3) \; = \; \frac 1{12} \, + \, \sqrt{C}
\hspace{10pt} .
$$
In the case $\Lambda > 0$ the scale factor $a(\eta)$ has a simple pole
at $\eta=\eta_\infty$ with residue $-\sqrt{3/\Lambda}$, i.\,e.\
$$
a(\eta) \; = \; \frac{\sqrt{3/\Lambda}}{\eta_\infty-\eta} \; - \;
\frac 3 \Lambda B \; + \; O(\eta-\eta_\infty)
\hspace{10pt} ,
$$
which leads to an exponential expansion of the universe with scale
factor $R(t) = a(\eta(t)) = O( \exp( \sqrt{\Lambda/3}\, t ) )$ for
cosmic time $t\to \infty$.

For $\Lambda=0$ and $\Omega_{\text{x}}>0$ one has at
$\eta=\eta_\infty$ a double pole, i.\,e.\
$$
a(\eta) \; = \; \frac{1/B}{(\eta-\eta_\infty)^2} \; - \; \frac 1{12B}
\; + \; O((\eta-\eta_\infty)^2)
\hspace{10pt} ,
$$
leading to $R(t) = B t^2 + \dots$ for $t\to \infty$, which follows
from the exact formula
$$
t \; = \; - \, \frac 1{12 B} \, \eta(t) \, + \, \frac 1B
\zeta(\eta_\infty-\eta(t)) \, - \, \frac 1B \zeta(\eta_\infty)
\hspace{10pt} ,
$$
where $\zeta(\eta):=\zeta(\eta;g_2,g_3)$ denotes the Weierstrass zeta
function.  In this special case formula (\ref{Eq:Scale_Factor})
reduces to (${\cal P}(\eta_\infty) = \frac 1{12}$)
$$
a(\eta) \; = \; \frac 1B {\cal P}(\eta-\eta_\infty) \, - \, \frac
1{12B}
\hspace{10pt} \hbox{ with } \hspace{10pt} 0 \leq \eta < \eta_\infty
\hspace{10pt} .
$$

Finally, if both the cosmological constant and the $X$-component vanish,
$B=C=0$, the invariants simplify to $g_2=\frac 1{12}$ and
$g_3=-\frac 1{216}$, and the ${\cal P}$-function can be
expressed in terms of an elementary function
\cite{Abramowitz_Stegun_1965}
$$
{\cal P}\left(\eta; \frac 1{12},-\frac 1{216}\right) \; = \; \frac
1{12} \, + \, \frac 1{4\sinh^2 \frac \eta 2}
\hspace{10pt} ,
$$
which, with (\ref{Eq:Scale_Factor}), leads to
$
a(\eta) = A ( \hat\eta \sinh \eta + \cosh \eta - 1 ),
$
which is the well-known expression for the scale factor of a
two-component model consisting of radiation and matter only.  This
leads to $R(t) = t + A\ln t + O(1)$ for $t\to\infty$.

As an illustration we show in figure \ref{Fig:scale_factor}
the cosmic scale factor $a(\eta)$ for three different, nearly flat
models ($\Omega_{\text{tot}} = 0.9$).
The full curve corresponds to a two-component model consisting of
radiation and matter ($\Omega_{\text{mat}} = 0.9$) only.
For this model $\eta_\infty=\infty$ holds.
The dashed curve represents a three-component model consisting of
radiation, matter and an $X$-component ($\Omega_{\text{x}} = 0.6$).
The approach to the double pole at $\eta_\infty=1.739$ is clearly visible.
The dotted curve shows $a(\eta)$ for a three-component model consisting of
radiation, matter and vacuum energy ($\Omega_{\text{vac}} = 0.6$).
One observes a steep rise to the pole at $\eta_\infty=1.385$.
In addition we have indicated the present scale factor $a_0$ by a dot.

In table \ref{Tabelle} the scale factor $a_0$ and the present age of
the universe $t_0$ as well as several cosmologically important times,
i.\,e.\ $\eta_{\text{eq}}$, $\eta_{\hbox{\scriptsize{SLS}}}$, $\eta_0$
and $\eta_\infty$, are given for several combinations of $\Omega_k$.
The present age of the universe $t_0$ is very close to the limit given
by globular cluster ages $13.5\pm 2.0$ Gyr \cite{Jimenez_1998,Primack_2000}.
White dwarf cooling rates lead to an age of our galaxy of $9.3\pm 2.0$ Gyr
\cite{Winget_Hansen_Liebert_Horn_Fontaine_Nather_Kepler_Lamb_1987}
or $8.0\pm 1.5$ Gyr \cite{Leggett_Ruiz_Bergeron_1998}. 
For a smaller Hubble constant, e.\,g.\ $h_0=0.6$, the age $t_0$
obtains larger values with 13.18 Gyr $\leq t_0 \leq$ 15.24 Gyr.

\section{The CMB anisotropy}
\label{computation_of_CMB}

In the following we consider only scalar perturbations and their
influence on the CMB.
Furthermore, we assume that the vacuum energy $\varepsilon_{\text{vac}}$
and the other dark energy component $\varepsilon_{\text{x}}$
are spatially constant.
The metric with scalar perturbations is written
in the conformal-Newtonian gauge in terms of scalar functions $\Phi$ and
$\Psi$ as
$$
ds^2 \; = \; a^2(\eta) \left\{ (1+2\Phi) d\eta^2 - (1-2\Psi)
\gamma_{ij} dx^i dx^j \right\}
\hspace{10pt} ,
$$
where $\Phi=\Psi$ for a diagonal $T_{\mu\nu}$.  Assuming negligible
entropy perturbations $\delta S=0$, the evolution of the metric
perturbation $\Phi$ gives in first-order perturbation theory in the
conformal-Newtonian gauge \cite{Mukhanov_Feldman_Brandenberger_1992}
$$
\Phi'' + 3 \hat H (1+c_s^2) \Phi' \, - \, c_s^2 \Delta \Phi
\hspace{120pt}
$$
$$ \hspace{60pt} + \, \{2 \hat H' + (1+3c_s^2)(\hat H^2+1)\} \Phi \; =
\; 0
\hspace{10pt} ,
$$
where $\hat H := a'/a$ and $\Delta$ denotes the Laplace-Beltrami operator.
The quantity $c_s^2$ can be interpreted as the sound velocity
and is given by
$$
c_s^2 \; = \;
\frac 1{3 + \frac 94 \varepsilon_{\text{mat}}/\varepsilon_{\text{rad}}}
\hspace{10pt} .
$$
Here it is assumed that the vacuum energy $\varepsilon_{\text{vac}}$
and the other dark energy component $\varepsilon_{\text{x}}$
are spatially constant.

Specifying $\Phi$ at $\eta=0$ such that it
corresponds to a scale-invariant (Harrison-Zel'dovich) spectrum,
allows the computation of the time-evolution of the metric
perturbation $\Phi$.  This in turn gives the input to the Sachs-Wolfe
formula \cite{Sachs_Wolfe_1967} which reads for isentropic initial
conditions
\begin{eqnarray}
\hspace{20pt} \frac{\delta T}T & = & \nonumber 2
\Phi(\eta_{\hbox{\scriptsize{SLS}}},\vec
x(\eta_{\hbox{\scriptsize{SLS}}})) - \, \frac 32 \Phi(0,\vec x(0)) \\
\label{Eq:Sachs_Wolfe} & & \hspace{50pt} + 2 \, \int_{\vec
x(\eta_{\hbox{\scriptsize{SLS}}})}^{\vec x(\eta_0)} d\eta
\frac{\partial\Phi(\eta,\vec x(\eta))}{\partial\eta}
\hspace{10pt} ,
\end{eqnarray}
from which one obtains the desired temperature fluctuations $\delta T$
of the CMB.
The conformal time at recombination,
which defines the surface of last scattering, is denoted by
$\eta_{\hbox{\scriptsize{SLS}}}$.  For $\eta_{\hbox{\scriptsize{SLS}}}
\gg \eta_{\hbox{\scriptsize{eq}}}$ the first two terms on the
right-hand side are approximately
\begin{equation}
\label{Eq:NSW_approx}
2 \Phi(\eta_{\hbox{\scriptsize{SLS}}},\vec
x(\eta_{\hbox{\scriptsize{SLS}}})) - \, \frac 32 \Phi(0,\vec x(0)) \;
\simeq \; \frac 13 \Phi(\eta_{\hbox{\scriptsize{SLS}}},\vec
x(\eta_{\hbox{\scriptsize{SLS}}}))
\hspace{10pt} .
\end{equation}
This is the so-called ordinary or naive Sachs-Wolfe term (NSW),
whereas the other term in (\ref{Eq:Sachs_Wolfe}) is called integrated
Sachs-Wolfe term (ISW).

\begin{figure}
\begin{center}
\vspace*{-10pt}
\hspace*{-1cm}\begin{minipage}{8.5cm}
\begin{minipage}{8.5cm}
\includegraphics[width=8.5cm]{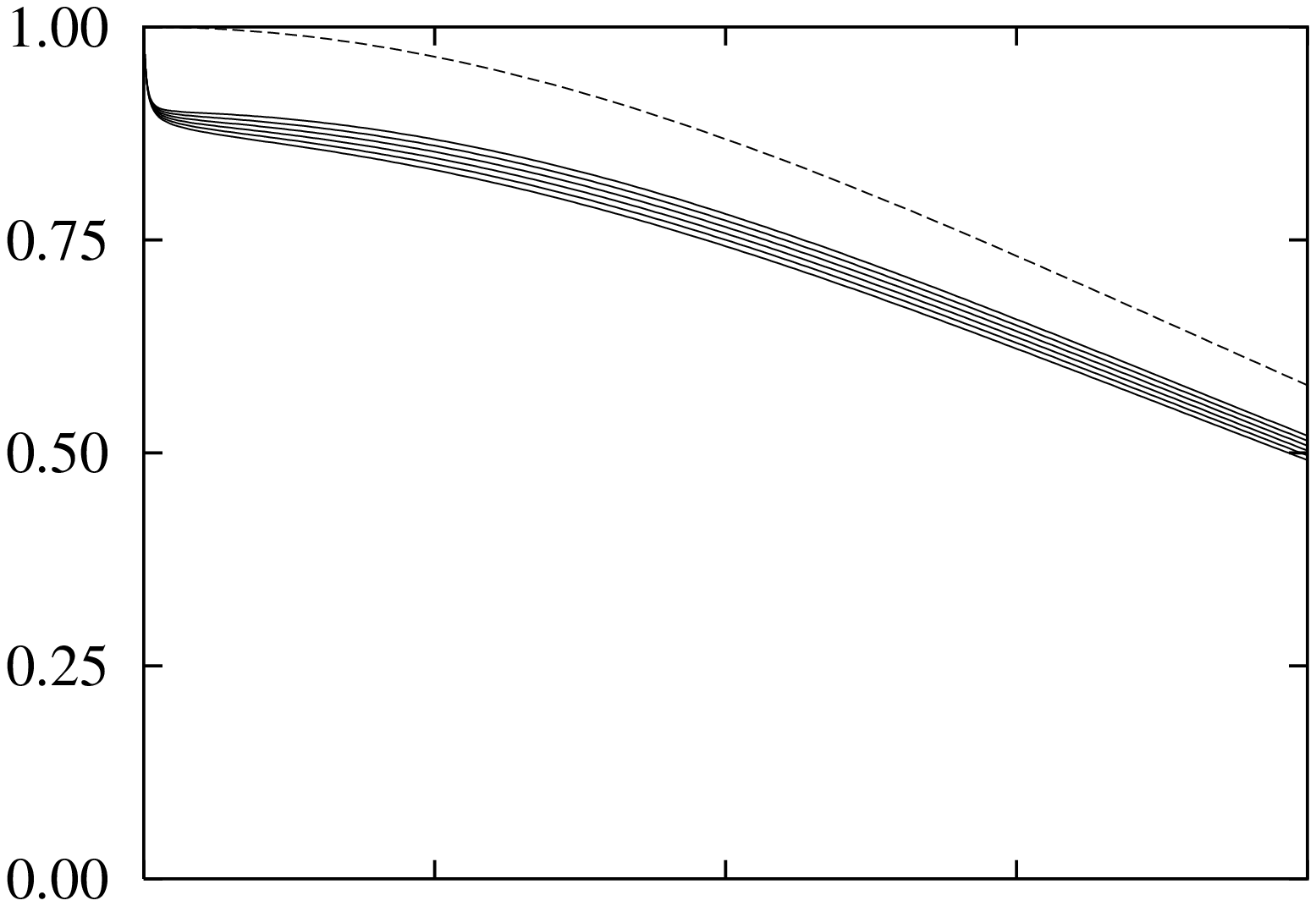}
\put(-30,143){a)} \put(-180,70){$\Omega_{\text{mat}} = 0.9$}
\put(-180,55){$\Omega_{\text{x}} = 0.0$}
\put(-180,40){$\Omega_{\text{vac}} = 0.0$}
\end{minipage}
\begin{minipage}{8.5cm}
\vspace*{-35pt}
\includegraphics[width=8.5cm]{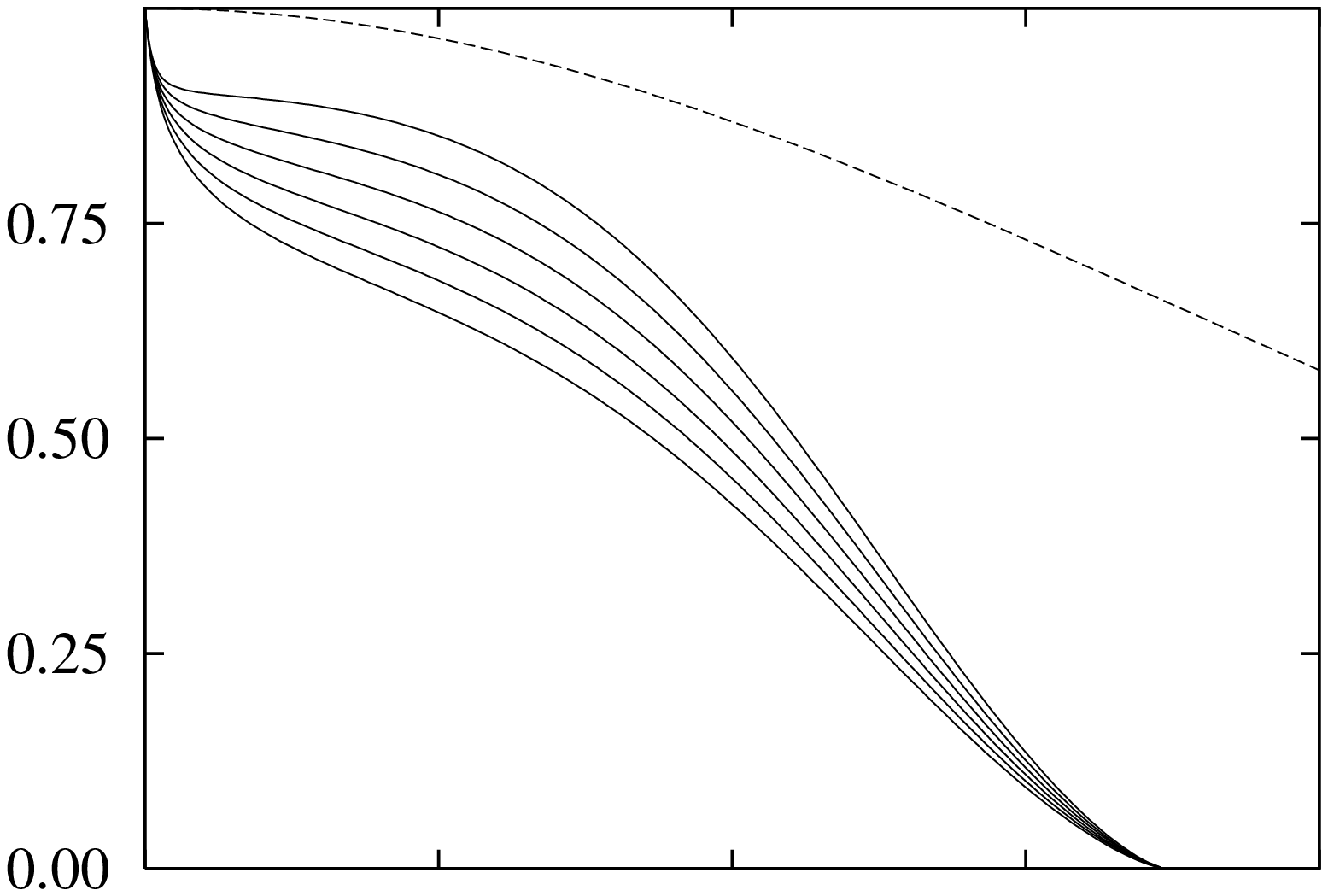}
\put(-30,143){b)} \put(-180,70){$\Omega_{\text{mat}} = 0.3$}
\put(-180,55){$\Omega_{\text{x}} = 0.6$}
\put(-180,40){$\Omega_{\text{vac}} = 0.0$}
\end{minipage}
\begin{minipage}{8.5cm}
\vspace*{-35pt}
\includegraphics[width=8.5cm]{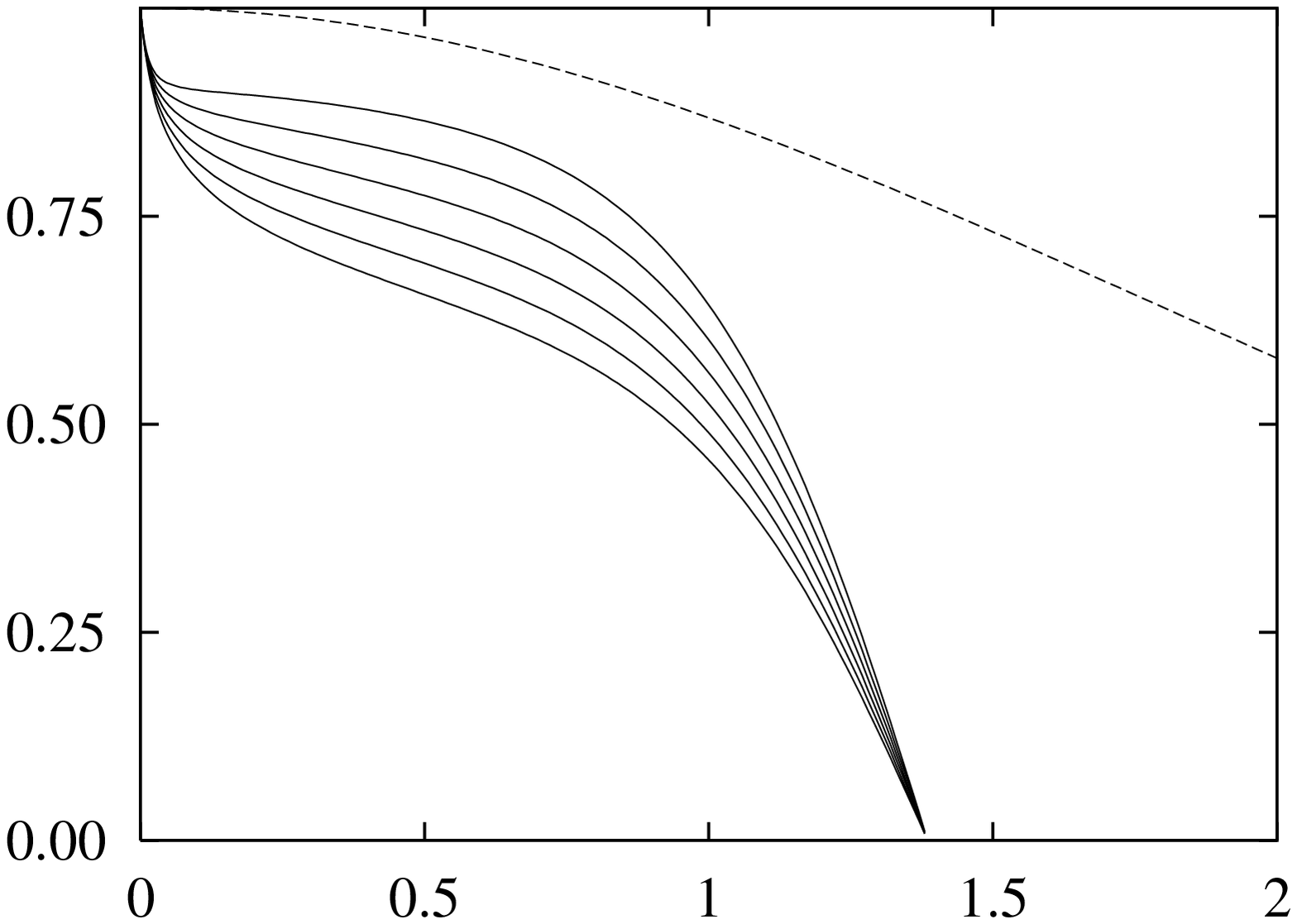}
\put(-30,143){c)} \put(-22,12){$\eta$}
\put(-180,70){$\Omega_{\text{mat}} = 0.3$}
\put(-180,55){$\Omega_{\text{x}} = 0.0$}
\put(-180,40){$\Omega_{\text{vac}} = 0.6$}
\end{minipage}
\end{minipage}
\end{center}
\vspace*{-10pt}
\caption{\label{Fig:f_n} The dependence of $f_n(\eta)/f_n(0)$ on the
eigenvalue $E_n$ is shown for $\Omega_{\text{tot}}=0.9$ and $h_0=0.7$.
The upper full curve corresponds to $E_n=0$ and the lowest one to
$E_n=5000$.  For the intermediate curves the energy is increased in
steps of 1000.  The dashed curve represents the result
$f_{\hbox{\scriptsize{mat}}}(\eta) =
5(\sinh^2\eta-3\eta\sinh\eta+4\cosh\eta-4) /(\cosh\eta-1)^3$ belonging
to a pure-matter model with $c_s=0$ used in some related works.  In a)
the matter dominated case $\Omega_{\text{mat}}=0.9$ is shown, whereas
the other two cases shown in b) and c) belong to
$\Omega_{\text{mat}}=0.3$ and $\Omega_{\text{x}}=0.6$,
$\Omega_{\text{vac}}=0.0$, respectively, $\Omega_{\text{x}}=0.0$,
$\Omega_{\text{vac}}=0.6$.  }
\end{figure}

The metric perturbation $\Phi$ is expanded with respect to the
eigenmodes
$$
\Delta \, \psi_n(\vec x\,) \; = \; -\, E_n \psi_n(\vec x\,)
\hspace{10pt} , \hspace{10pt} k_n := \sqrt{E_n-1}
\hspace{10pt} ,
$$
of the considered compact orbifold, i.\,e.\
$$
\Phi(\eta,\vec x\,) \; = \; \sum_{n=1}^\infty f_n(\eta)\, \psi_n(\vec
x\,)
\hspace{10pt} ,
$$
which yields for $f_n(\eta)$ the differential equation
\begin{eqnarray}\nonumber
\hspace{10pt} f_n''(\eta) & + & 3 \hat H (1+c_s^2) f_n'(\eta) \\ & &
\hspace{-25pt} \label{Eq:evolution_f} + \, \{ c_s^2 E_n + 2 \hat H' +
(1+3c_s^2)(\hat H^2+1)\} f_n(\eta) \; = \; 0
\hspace{5pt} , \hspace{5pt}
\end{eqnarray}
where $\hat{H}$ and $c_s^2$ are determined by the background model.
The initial conditions are ($\alpha>0$ is a normalization constant)
\begin{equation}
\label{Eq:Harrison_Zeldovitch}
f_n(0) \; = \; \frac{\alpha}{\sqrt{k_n(k_n^2+1)}}
\hspace{10pt} \hbox{ and } \hspace{10pt} f_n'(0) \; = \; - \,
\frac{f_n(0)}{8\hat{\eta}}
\hspace{10pt} ,
\end{equation}
which carry over to a Harrison-Zel'dovitch spectrum having a spectral
index $n=1$ and selecting only the non-decaying modes.
Using the eigenmodes the perturbation is defined obeying the periodicity
condition imposed by the fundamental cell.

The time dependence of $f_n(\eta)$, determined by the background model
(\ref{Eq:Scale_Factor}), is obtained by numerical integration of
(\ref{Eq:evolution_f}) and is shown in figure \ref{Fig:f_n} for three
different models.
The first model, shown in figure \ref{Fig:f_n}a), is completely dominated by
matter, $\Omega_{\text{tot}}=\Omega_{\text{mat}}=0.9$, whereas the other two
models belong to $\Omega_{\text{mat}}=0.3$ and 0.6 for
$\Omega_{\text{x}}$ and $\Omega_{\text{vac}}$, shown in b) and c),
respectively.  The latter two cases have a finite $\eta_\infty$ at
which the perturbation vanishes, i.\,e.\ $(\eta\to\eta_\infty)$
$$
f_n(\eta) \; \propto \; (\eta_\infty-\eta)^{\frac{7-\sqrt{17}}2}
$$
for the case $\Omega_{\text{x}}>0$ and $\Omega_{\text{vac}}=0$,
and
$$
f_n(\eta) \; \propto \; \eta_\infty-\eta
$$
for $\Omega_{\text{x}}=0$ and $\Omega_{\text{vac}}>0$.
Furthermore, perturbation modes with wavelength $\lambda=\frac{2\pi}k
> \eta_\infty$ will never enter the horizon in models with
$\Omega_{\text{x}}>0$ and/or $\Omega_{\text{vac}}>0$.  The first
decline of $f_n(\eta)/f_n(0)$ from 1 to $\frac 9{10}$ for small values of
$\eta$ is due to the transition from the radiation- to the
matter-dominated epoch (see, e.\,g.\
\cite{Mukhanov_Feldman_Brandenberger_1992}), which leads to the
approximation (\ref{Eq:NSW_approx}).

\begin{figure}
\vspace*{-15pt}
\begin{center}
\hspace*{-25pt}\includegraphics[width=9cm]{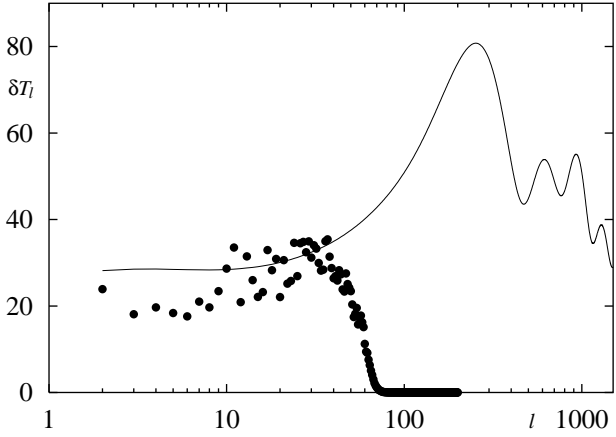}
\put(-38,12){$l$} \put(-235,138){$\delta T_l$}
\end{center}
\vspace*{-15pt}
\caption{\label{Fig:Cl_CMBFast}
The angular power spectrum
$\delta T_l = \sqrt{l(l+1)C_l/2\pi}$ is shown for a model with
$\Omega_{\text{bar}}=0.05$, $\Omega_{\text{cdm}}=0.25$,
$\Omega_{\text{x}}=0.0$, $\Omega_{\text{vac}}=0.6$
and $h_0=0.7$.
The curve is obtained from CMBFAST, whereas the dots show the result
by using the discrete eigenmode spectrum.
}
\end{figure}

With the background model (\ref{Eq:Scale_Factor}), the time-evolution
(\ref{Eq:evolution_f}), and the Sachs-Wolfe formula
(\ref{Eq:Sachs_Wolfe}), the angular power spectrum of the CMB
anisotropy
$$
C_l \; = \; \frac{1}{2l+1}\,\sum_{m=-l}^l |a_{lm}|^2
$$
can be computed, where $a_{lm}$ are the expansion coefficients of the
CMB anisotropy $\delta T$ with respect to the spherical harmonics
$Y_l^m(\theta,\phi)$.

The angular power spectra $\delta T_l := \sqrt{l(l+1)C_l/2\pi}$
are computed for several models.
The considered compact orbifold as well as the position
of the observer is the same as in \cite{Aurich_1999}.
In figure \ref{Fig:Cl_CMBFast} the angular power spectrum $\delta T_l$
is shown for the case $\Omega_{\text{bar}}=0.05$, $\Omega_{\text{cdm}}=0.25$,
$\Omega_{\text{x}}=0.0$, $\Omega_{\text{vac}}=0.6$
and $h_0=0.7$.
The curve is obtained from CMBFAST \cite{Zaldarriaga_Seljak_1999},
i.\,e.\ is obtained for an infinitely extended hyperbolic universe,
whereas the dots are obtained by the procedure outlined above,
i.\,e.\ for a compact hyperbolic universe.
Because the latter computation takes only the first 749 eigenmodes with
$k<55$ into account, one observes at $l \sim 40$ a decline towards zero.
This is solely due to the truncation in the $k$-summation
because the eigenmodes are only computed up to this $k$ value.
The considered modes are all above horizon at recombination,
and thus the processes leading to the acoustic peak can be ignored.
In the following we are only concerned with the low multipoles $C_l$
which are affected by the non-trivial topology.
These low multipoles are not affected by truncating the $k$-summation.
The figure shows clearly the suppression in power for $l\lesssim 10$.

Since $\delta T_l$ is computed for a fixed observer,
i.\,e. it represents a ``one-sky realization'', the spectrum is not
smooth like the CMBFAST spectrum.
It rather shows fluctuations which an experiment would observe
which necessarily measures a one-sky realization.

\begin{figure}
\vspace*{-15pt}
\hspace*{-10cm}
\begin{center}
\begin{minipage}{8cm}
\includegraphics[width=8cm]{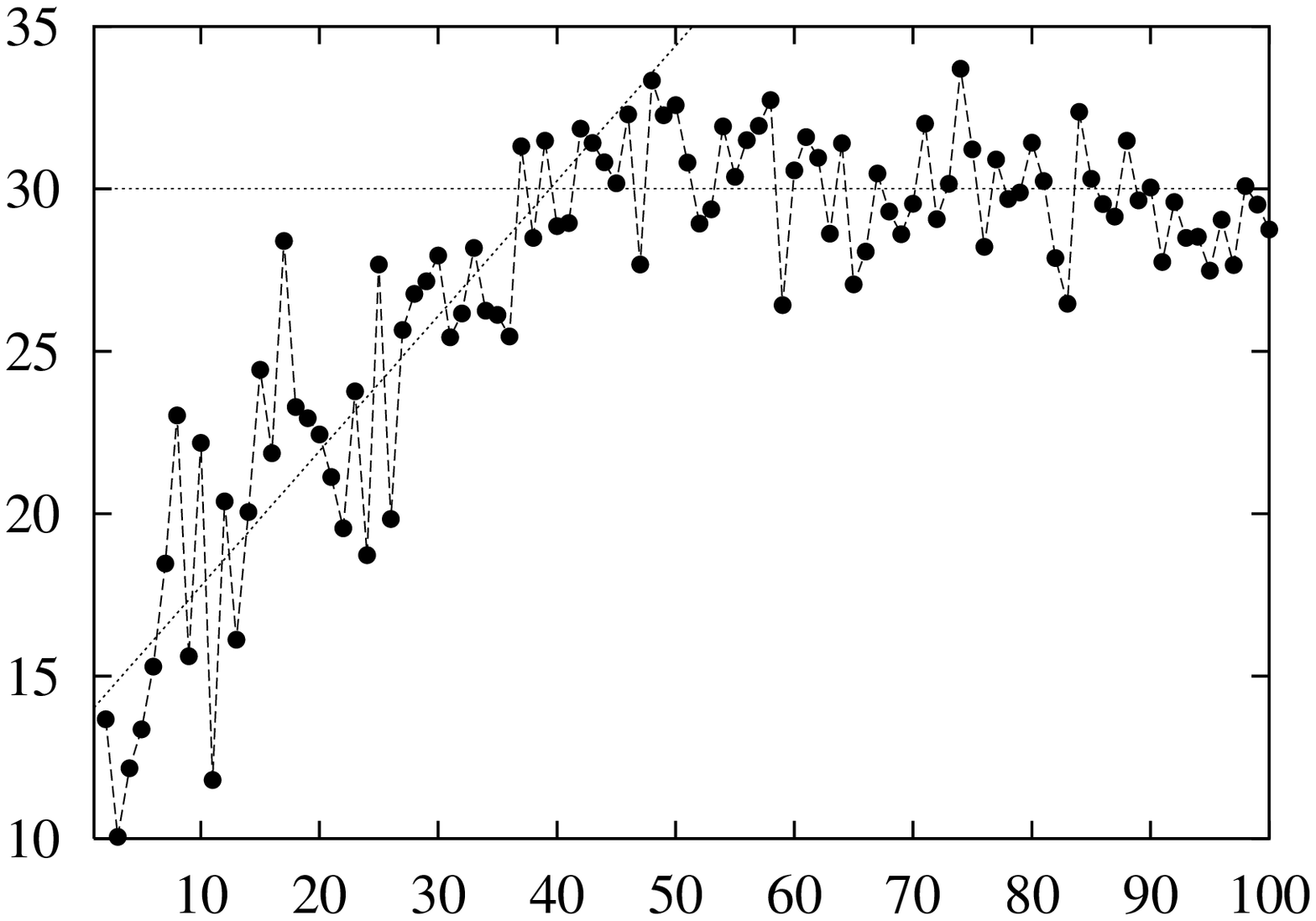}
\put(-187,137){a)} \put(-211,137){$\delta T_l$} \put(-17,10){$l$}
\put(-70,65){$\Omega_{\text{mat}}=0.3$}
\put(-70,50){$\Omega_{\text{x}}=0.0$}
\put(-70,35){$\Omega_{\text{vac}}=0.2$}
\end{minipage}
\vspace*{-10pt}
\begin{minipage}{8cm}
\includegraphics[width=8cm]{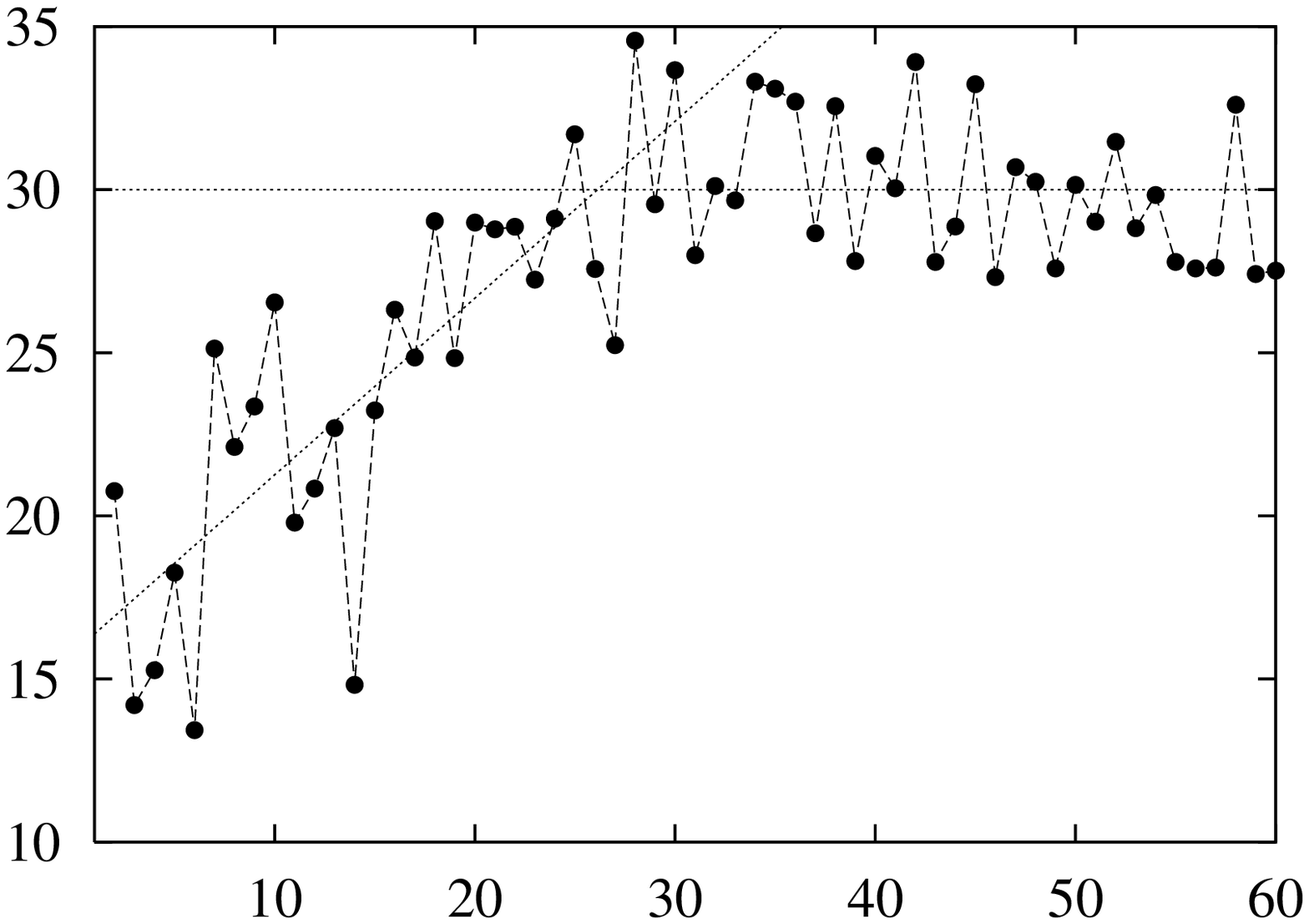}
\put(-187,137){b)} \put(-211,137){$\delta T_l$} \put(-25,10){$l$}
\put(-70,65){$\Omega_{\text{mat}}=0.3$}
\put(-70,50){$\Omega_{\text{x}}=0.0$}
\put(-70,35){$\Omega_{\text{vac}}=0.5$}
\end{minipage}
\begin{minipage}{8cm}
\includegraphics[width=8cm]{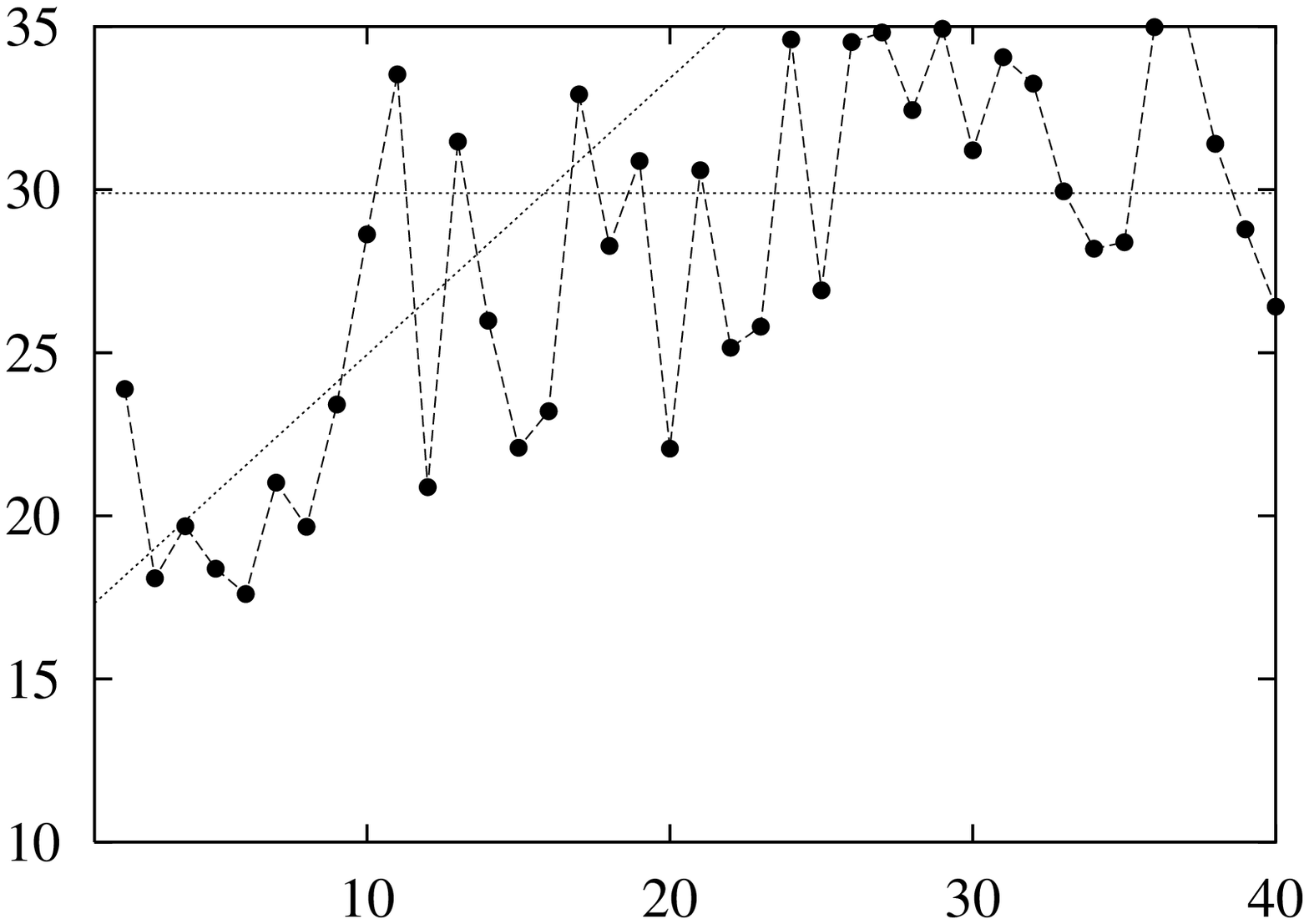}
\put(-187,137){c)} \put(-211,137){$\delta T_l$} \put(-25,10){$l$}
\put(-70,65){$\Omega_{\text{mat}}=0.3$}
\put(-70,50){$\Omega_{\text{x}}=0.0$}
\put(-70,35){$\Omega_{\text{vac}}=0.6$}
\end{minipage}
\begin{minipage}{8cm}
\includegraphics[width=8cm]{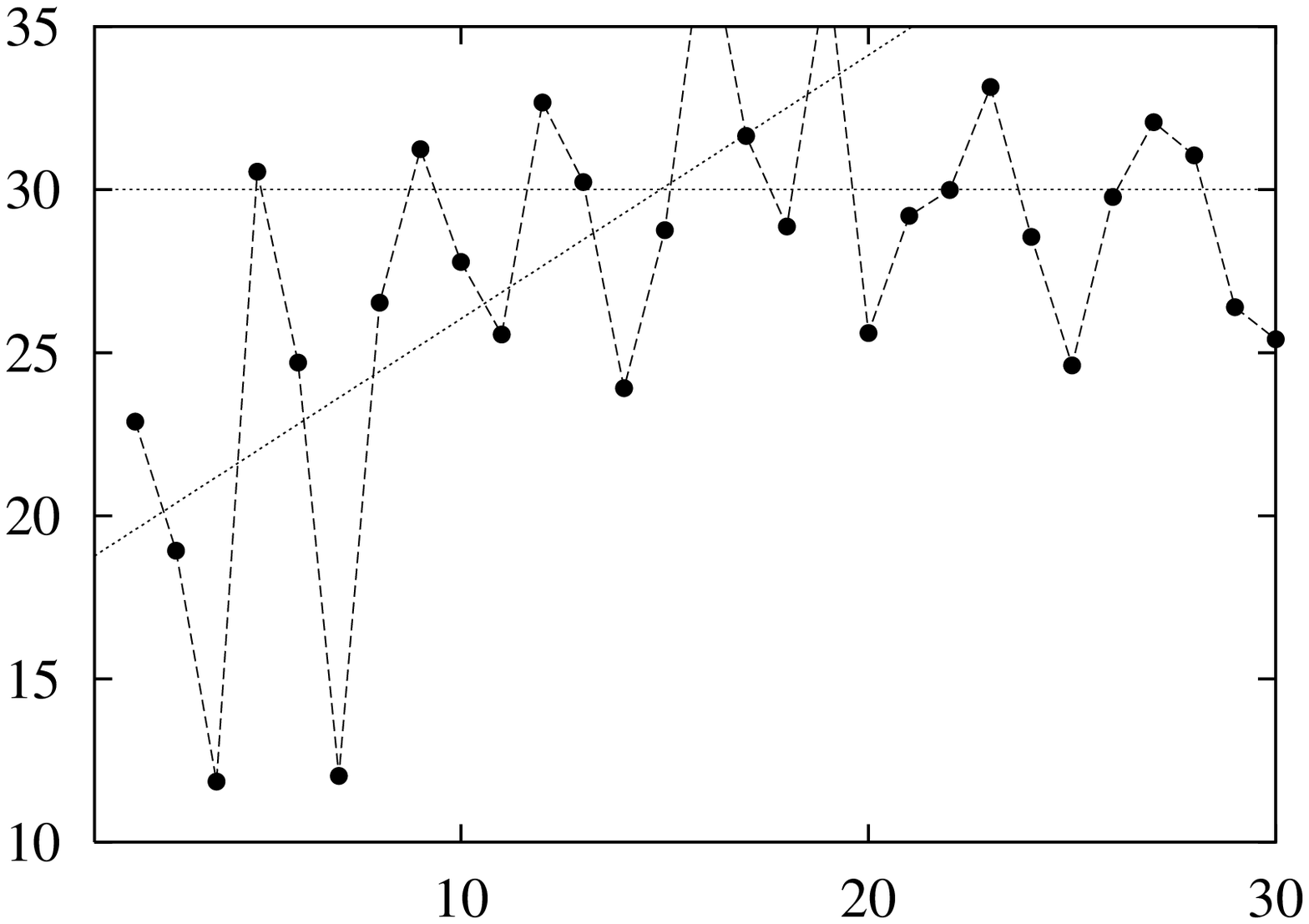}
\put(-187,137){d)} \put(-211,137){$\delta T_l$} \put(-25,10){$l$}
\put(-70,65){$\Omega_{\text{mat}}=0.3$}
\put(-70,50){$\Omega_{\text{x}}=0.0$}
\put(-70,35){$\Omega_{\text{vac}}=0.65$}
\end{minipage}
\end{center}
\vspace*{-10pt}
\caption{\label{Fig:Cl_lambda_Knick} The angular power spectrum
$\delta T_l = \sqrt{l(l+1)C_l/2\pi}$ is shown in $\mu K$ for models
with a vanishing $X$-component.  }
\end{figure}

\begin{figure}
\vspace*{-15pt}
\hspace*{-10cm}
\begin{center}
\begin{minipage}{8cm}
\includegraphics[width=8cm]{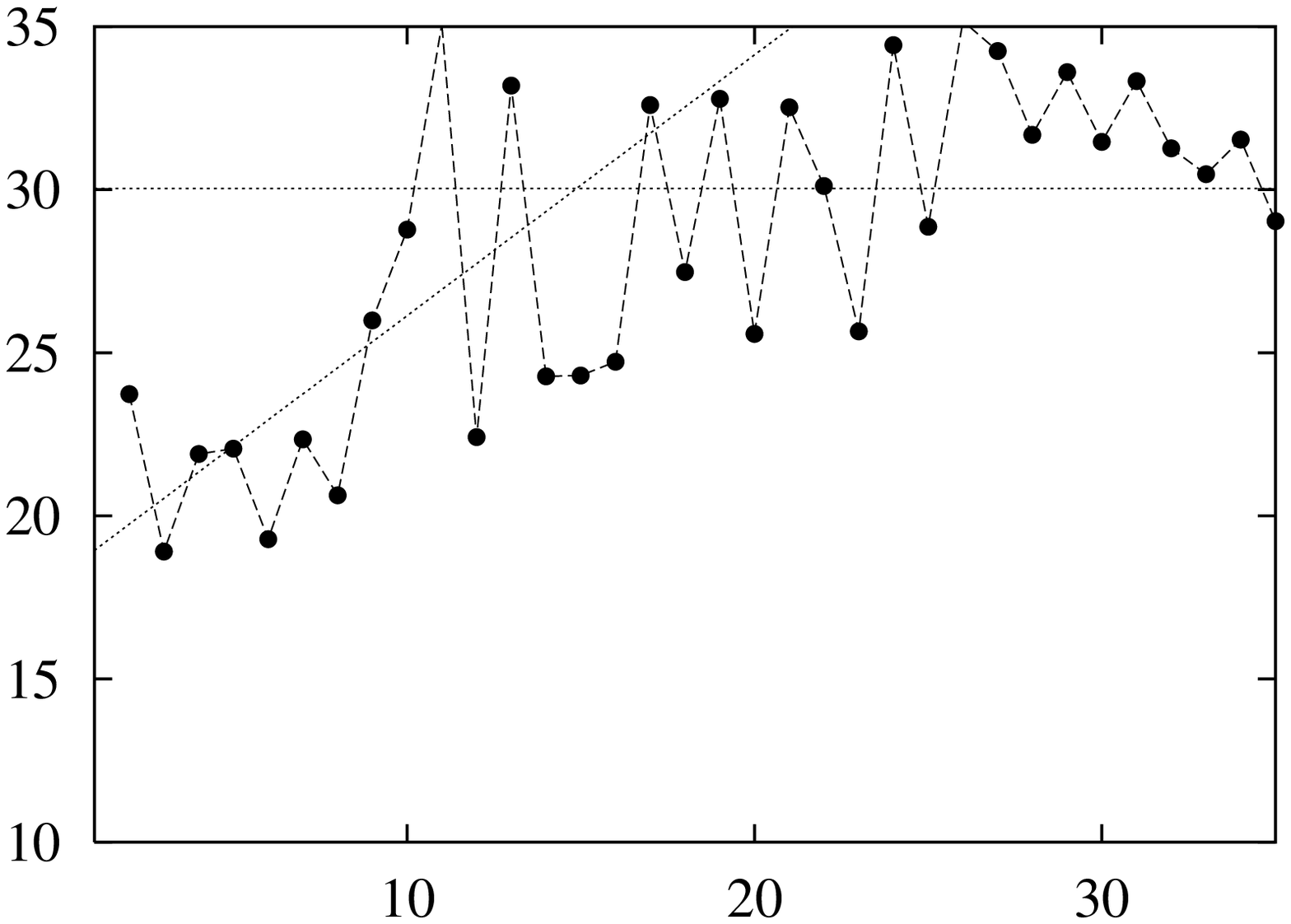}
\put(-187,137){a)} \put(-211,137){$\delta T_l$} \put(-17,10){$l$}
\put(-70,65){$\Omega_{\text{mat}}=0.3$}
\put(-70,50){$\Omega_{\text{x}}=0.3$}
\put(-70,35){$\Omega_{\text{vac}}=0.3$}
\end{minipage}
\vspace*{-10pt}
\begin{minipage}{8cm}
\includegraphics[width=8cm]{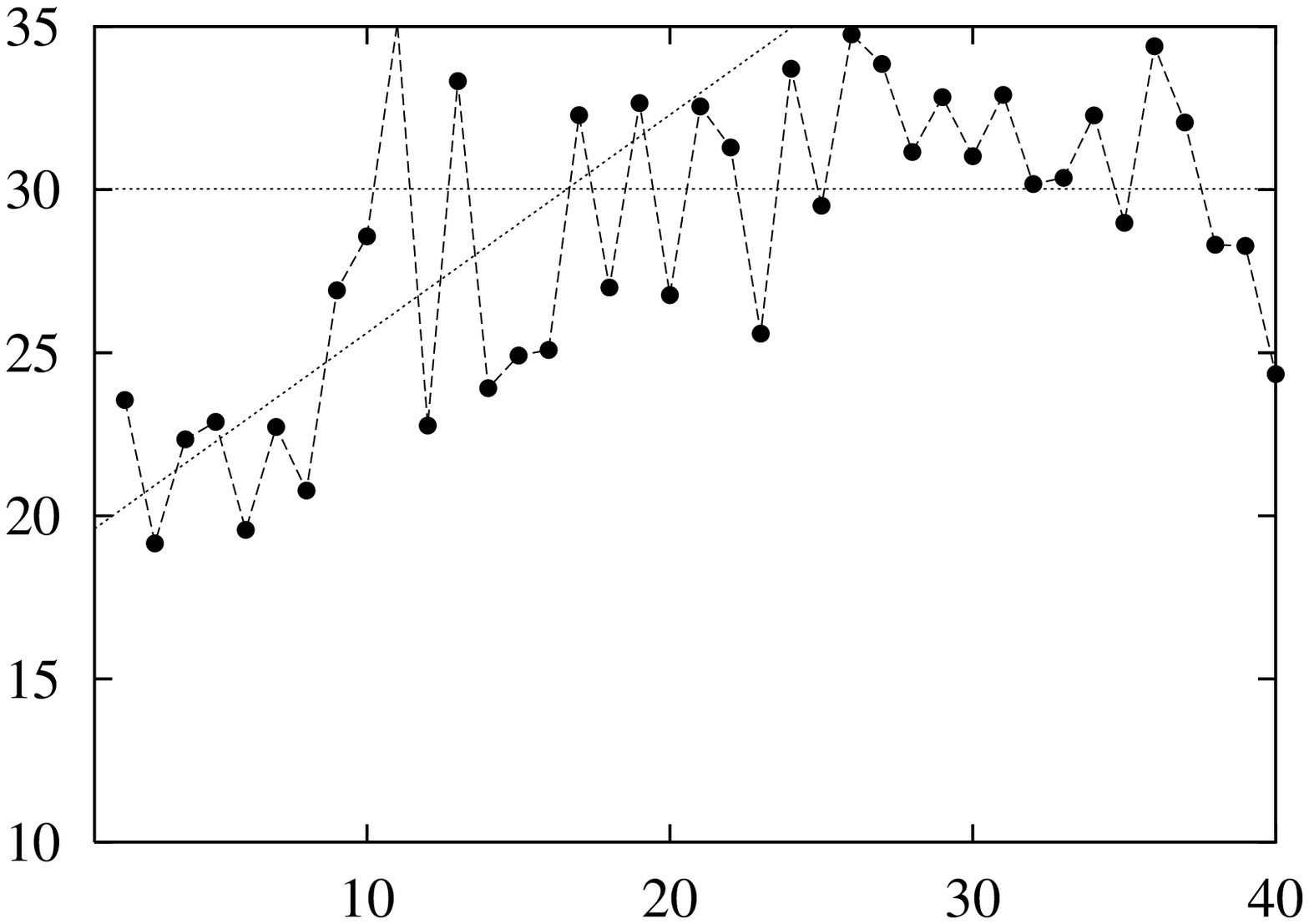}
\put(-187,137){b)} \put(-211,137){$\delta T_l$} \put(-25,10){$l$}
\put(-70,65){$\Omega_{\text{mat}}=0.3$}
\put(-70,50){$\Omega_{\text{x}}=0.4$}
\put(-70,35){$\Omega_{\text{vac}}=0.2$}
\end{minipage}
\begin{minipage}{8cm}
\includegraphics[width=8cm]{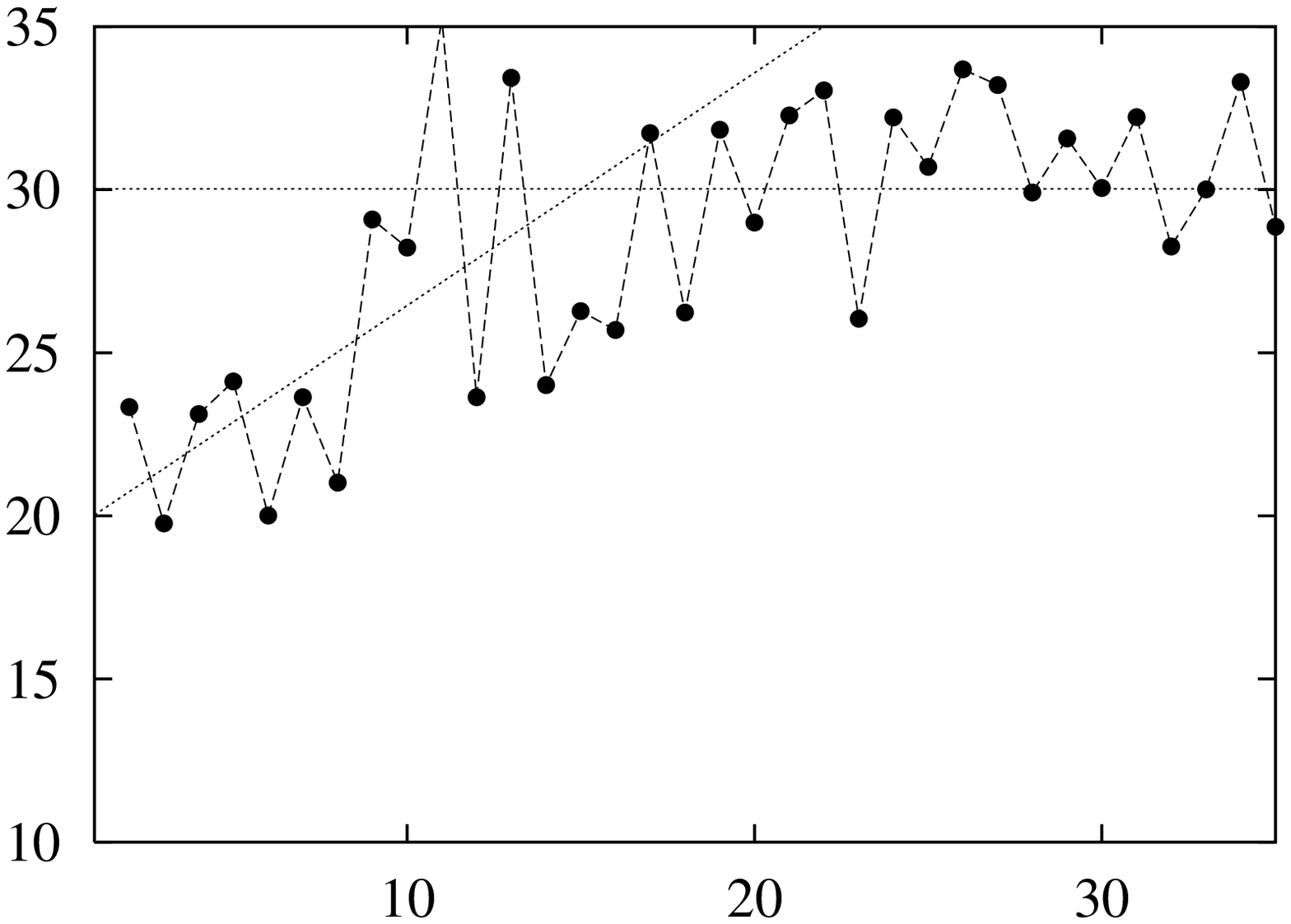}
\put(-187,137){c)} \put(-211,137){$\delta T_l$} \put(-25,10){$l$}
\put(-70,65){$\Omega_{\text{mat}}=0.3$}
\put(-70,50){$\Omega_{\text{x}}=0.6$}
\put(-70,35){$\Omega_{\text{vac}}=0.0$}
\end{minipage}
\begin{minipage}{8cm}
\includegraphics[width=8cm]{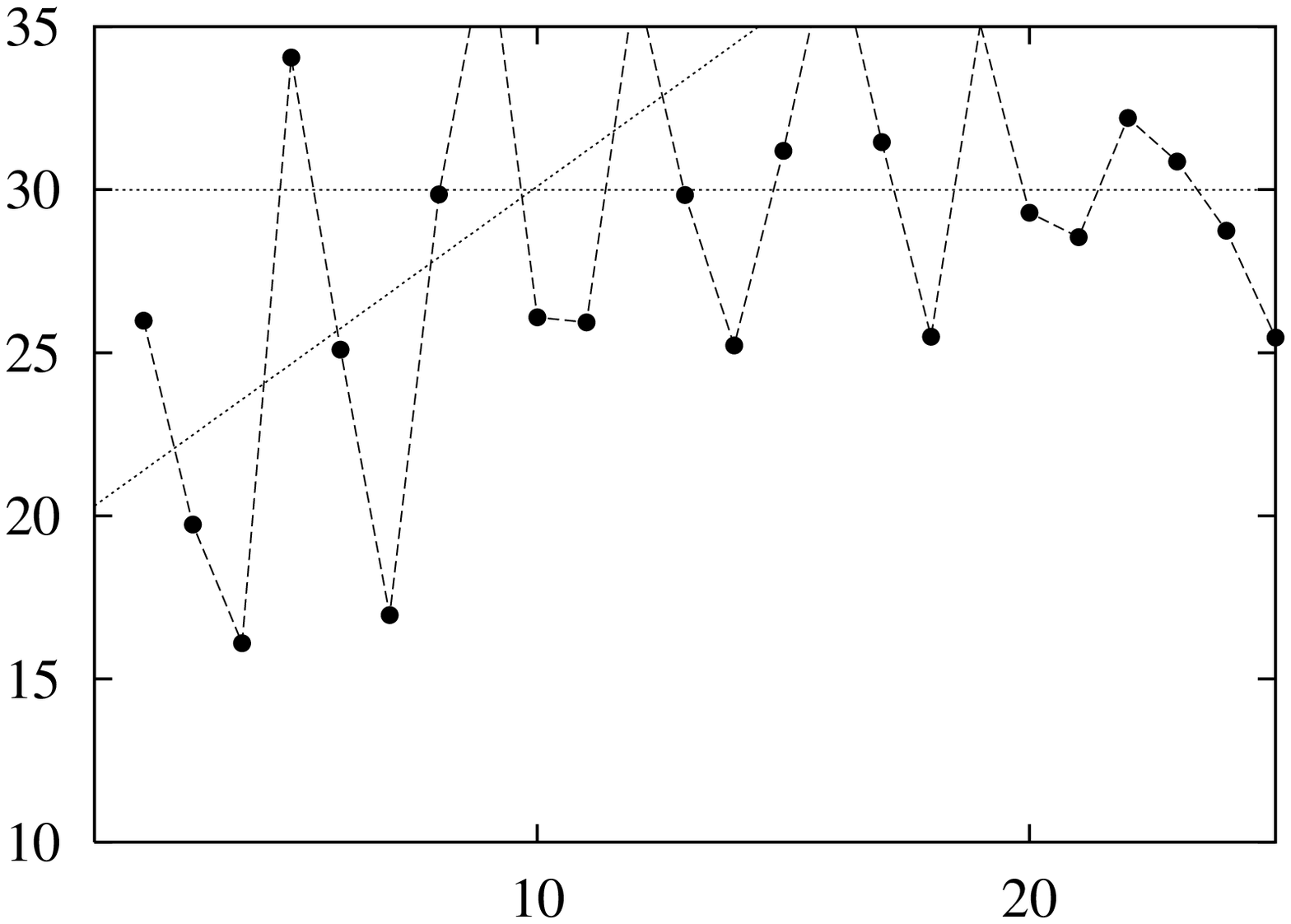}
\put(-187,137){d)} \put(-211,137){$\delta T_l$} \put(-25,10){$l$}
\put(-70,65){$\Omega_{\text{mat}}=0.3$}
\put(-70,50){$\Omega_{\text{x}}=0 .65$}
\put(-70,35){$\Omega_{\text{vac}}=0.0$}
\end{minipage}
\end{center}
\vspace*{-10pt}
\caption{\label{Fig:Cl_Quintessence_Knick} The angular power spectrum
$\delta T_l = \sqrt{l(l+1)C_l/2\pi}$ is shown in $\mu K$ for models
with an $X$-component.  }
\end{figure}

The lower part of the angular power spectra $\delta T_l$,
computed for several models with a vanishing and non-vanishing
$X$-component, are shown in figures \ref{Fig:Cl_lambda_Knick} and
\ref{Fig:Cl_Quintessence_Knick}, respectively.
The spectra in
figures \ref{Fig:Cl_lambda_Knick}a) and \ref{Fig:Cl_lambda_Knick}b)
for $\Omega_{\text{tot}}=0.5$ and $\Omega_{\text{tot}}=0.8$,
respectively, show a plateau being normalized to $30 \mu\hbox{K}$.  At
higher values of $l$ the angular power spectra $\delta T_l$ rise again
(see figure \ref{Fig:Cl_CMBFast}) which is not shown here because our
calculations do not take into account the necessary processes leading to
the acoustic peak, since the modes considered here are well above
the horizon at recombination.
For low values of $l$ one observes a nearly linear
increase of $\delta T_l$ which is caused by the finite size of the
fundamental cell which in turn causes a cut-off in the $k$-spectrum.
(The straight lines are drawn solely to guide the eyes.)  One observes
that the ``bend'' point, where the behavior turns from a linear
increase to a plateau, decreases towards smaller values of $l$
for increasing vacuum energy.
If the amount of vacuum energy is replaced by the same energy contribution
of an $X$-component one obtains quantitatively analogous
angular power spectra because the behavior of $f_n(\eta)$ shown in figure
\ref{Fig:f_n} is similar for vacuum energy and the $X$-component
for $\eta \lesssim \eta_0$.
The two models shown in figures \ref{Fig:Cl_lambda_Knick}c) and
\ref{Fig:Cl_lambda_Knick}d) possess an even larger $\Omega_{\text{tot}}$,
i.\,e.\ $\Omega_{\text{tot}}=0.9$ and $\Omega_{\text{tot}}=0.95$, respectively.
Here the suppression is much less pronounced than in the cases with
$\Omega_{\text{tot}} \lesssim 0.85$.
In the case $\Omega_{\text{tot}}=0.9$ the quadrupole moment is larger than
the other low multipoles which is due to the large integrated Sachs-Wolfe
contribution (see below).
In the other case $\Omega_{\text{tot}}=0.95$ one observes very large
fluctuations for low values of $l$.

For four models with an $X$-component the angular power spectra
$\delta T_l$ are shown in figure \ref{Fig:Cl_Quintessence_Knick}.
In figure \ref{Fig:Cl_Quintessence_Knick}a) and
\ref{Fig:Cl_Quintessence_Knick}b) two models with $\Omega_{\text{tot}}=0.9$
are shown, where in the first case the energy density is equally distributed
between $\Omega_{\text{mat}}$, $\Omega_{\text{x}}$ and $\Omega_{\text{vac}}$,
and in the second case the $X$-component dominates.
One observes similar angular power spectra which is again explained
by the similar behavior of $f_n(\eta)$.
In both cases the multipoles with $l\lesssim 10$ are suppressed. 
In figure \ref{Fig:Cl_Quintessence_Knick}c) and
\ref{Fig:Cl_Quintessence_Knick}d) two models with a vanishing
vacuum energy are shown for $\Omega_{\text{tot}}=0.9$ and
$\Omega_{\text{tot}}=0.95$, respectively.
In the latter case the suppression of low multipoles is blurred by
very large fluctuations, which occur as in figure \ref{Fig:Cl_lambda_Knick}d).


\onecolumn

\twocolumn

\begin{figure}
\hspace*{-25pt}\begin{minipage}{9cm}
\begin{center}
\includegraphics[width=9cm]{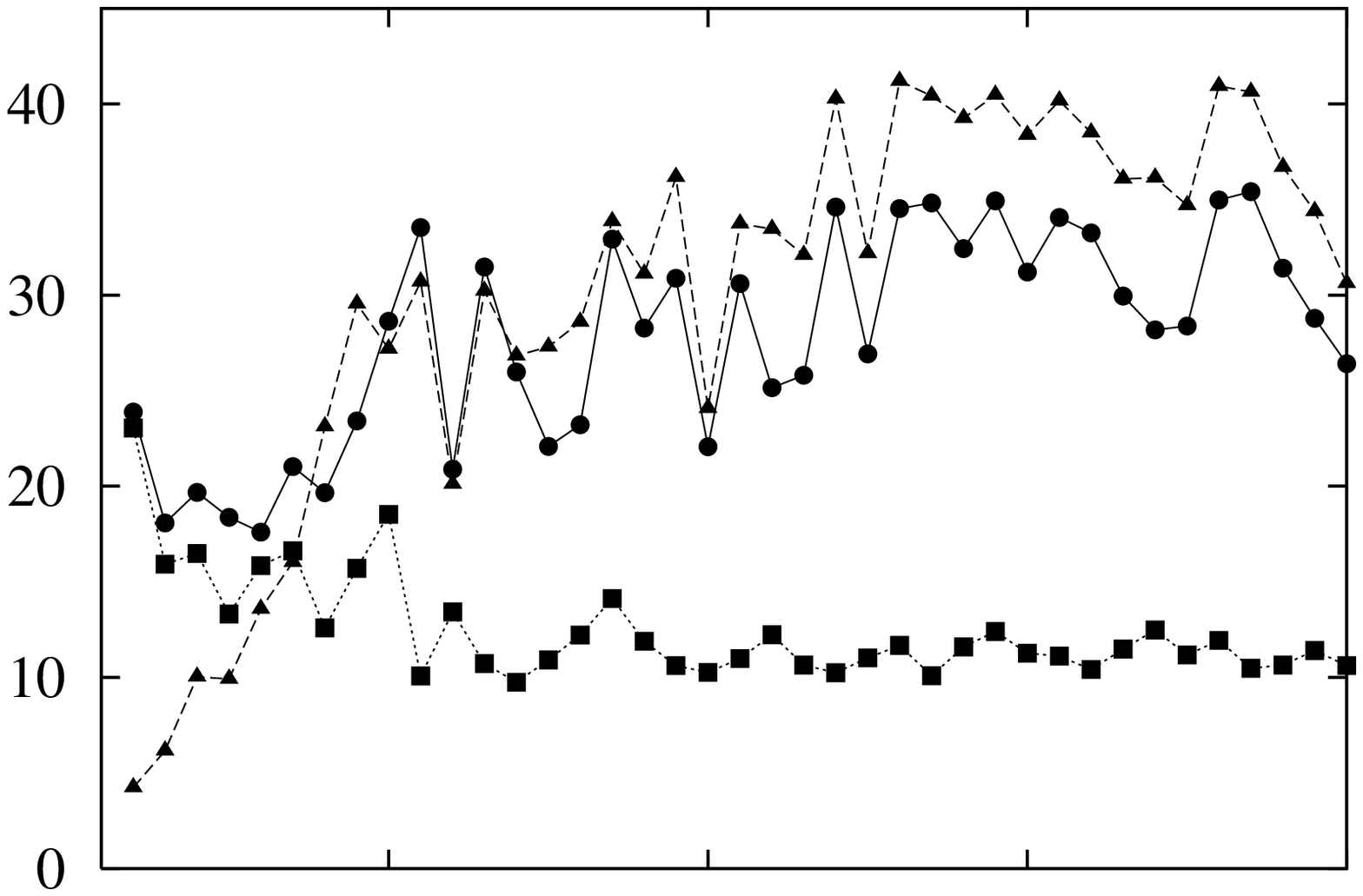}
\put(-235,140){$\delta T_l$}
\put(-200,150){a)}
\end{center}
\end{minipage}
\hspace*{-25pt}\begin{minipage}{9cm}
\vspace*{-36pt}
\begin{center}
\includegraphics[width=9cm]{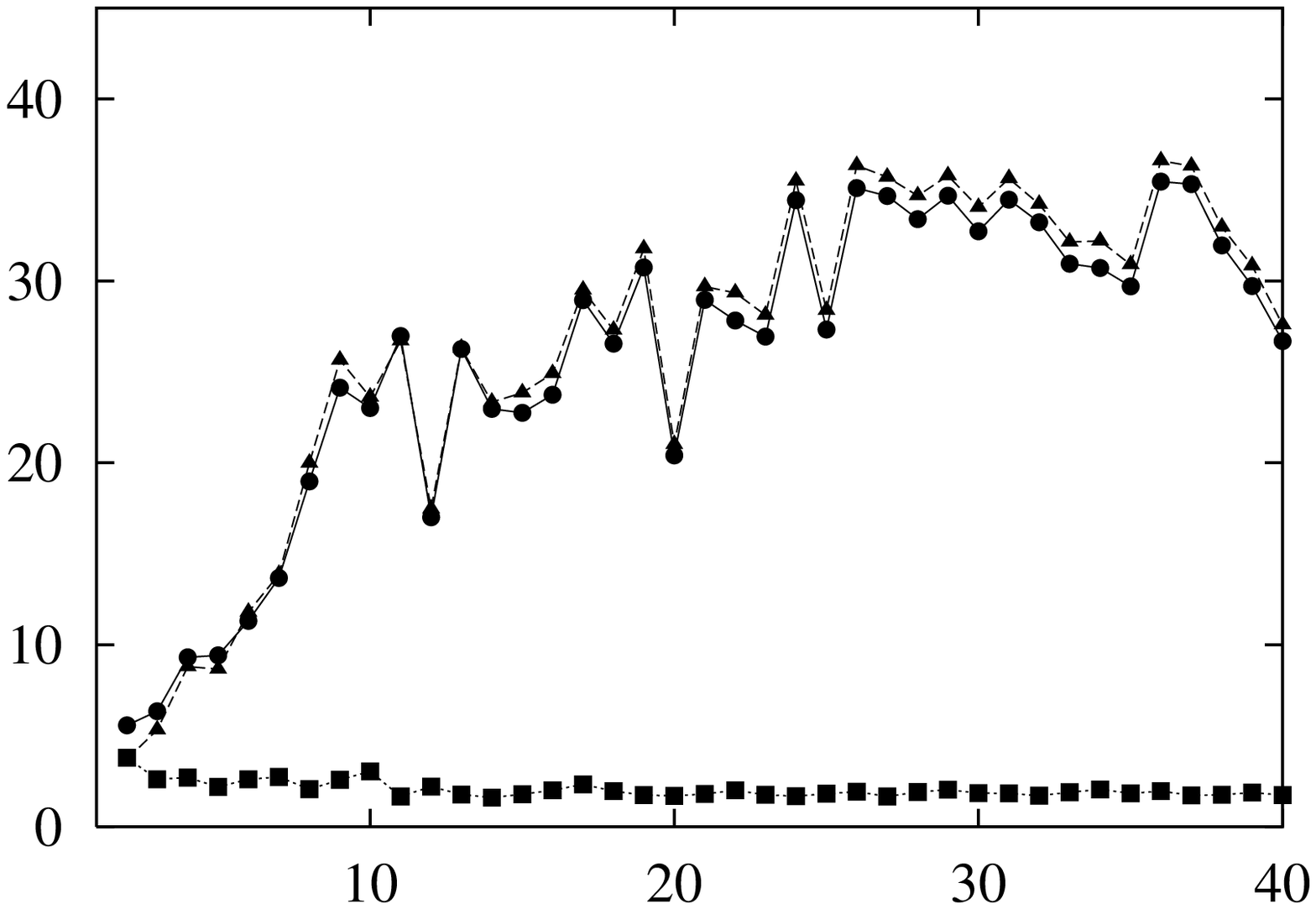}
\put(-30,12){$l$}
\put(-235,140){$\delta T_l$}
\put(-200,150){b)}
\end{center}
\end{minipage}
\vspace*{-15pt}
\caption{\label{Fig:NSW_ISW}
The angular power spectrum $\delta T_l = \sqrt{l(l+1)C_l/2\pi}$ is shown in
$\mu K$ with respect to the NSW- and ISW-contributions separately for the case
$\Omega_{\text{mat}} = 0.3$, 
$\Omega_{\text{x}} = 0.0$ and
$\Omega_{\text{vac}} = 0.6$.
In figure \protect\ref{Fig:NSW_ISW}a) isentropic initial conditions are chosen,
whereas in figure \protect\ref{Fig:NSW_ISW}b) isocurvature initial conditions
are assumed.
The full circles denote the sum of both contributions, and the triangles and
the squares correspond to the NSW- and ISW-contribution, respectively.
}
\end{figure}

The angular power spectrum does not go to zero at the smallest
values of $l$.
This is due to the competition of the two contributions to $\delta T_l$,
i.\,e.\ the NSW and the ISW term in (\ref{Eq:Sachs_Wolfe}).
As shown in figure \ref{Fig:NSW_ISW} the NSW term gives a contribution
which indeed vanishes for small values of $l$
because of the cut-off in the eigenmode spectrum.
But the ISW term adds an almost constant contribution
which even increases towards small values of $l$.
This interplay is responsible for the fact that $\delta T_l$
does not fall below $\sim 15\mu K$, where a plateau of $30\mu K$
is assumed.
The relative contribution of the two terms is largely determined
by the chosen initial conditions at $\eta=0$.

The inflationary models naturally suggest isentropic initial conditions
and these are imposed in the above calculations.
However, imposing isocurvature initial conditions leads to a much
smaller ISW contribution relative to the NSW term.
This is shown in figure \ref{Fig:NSW_ISW}, where in figure \ref{Fig:NSW_ISW}a)
isentropic initial conditions according to most inflationary models
and in figure \ref{Fig:NSW_ISW}b) isocurvature initial conditions are chosen.
If the observed increase in $\delta T_l$ would be approximately linear towards zero
for nearly flat models, this would imply a small ISW contribution
and this would point to isocurvature initial conditions.

To summarize the results, the anomalously low quadrupole moment obtained from
the COBE measurements can be taken as a first sign for a universe with a
finite volume.
The presented calculations demonstrate that low multipoles occur
for the considered compact fundamental domain even for nearly
flat, but hyperbolic, models with $\Omega_{\text{tot}} \lesssim 0.9$.
For even larger values of $\Omega_{\text{tot}} \simeq 0.95$
very large fluctuations occur which may also be an indication for
a finite volume.
Furthermore, the kind of increase of $\delta T_l$ gives a clue to the
initial conditions.
Future experiments which survey the complete CMB sky like MAP and PLANCK,
will have the required signal to noise ratio to reveal a possible finite universe.

\subsection*{ACKNOWLEDGMENTS}

We would like to thank the Rechenzentrum of the University of Karlsruhe for
the access to their computers.

\bibliography{bib_astro} 
\bibliographystyle{apalike}

\label{lastpage}

\end{document}